\documentclass[aps,preprintnumbers,amsmath,amssymb,prl,superscriptaddress,longbibliography,twocolumn]{revtex4-1}
\usepackage{dcolumn}
\usepackage{color}
\usepackage[caption=false]{subfig}
\usepackage{feynmp}
\usepackage{feynmp-auto}
\usepackage{systeme,mathtools}
\usepackage{easyReview}
\usepackage{physics}
\usepackage{dsfont}

\def\comment#1{}

\newcommand{\nc}{\newcommand}
\nc{\scs}{\scriptstyle}
\nc{\setval}{\fmfset{wiggly_len}{3mm} \fmfset{arrow_len}{1.5mm}
	\fmfset{arrow_ang}{13} \fmfset{dash_len}{1.5mm}\fmfpen{0.125mm}
	\fmfset{dot_size}{2thick}}

\usepackage{bm,latexsym,mathrsfs,enumerate,color}
\usepackage[mathcal]{euscript}
\usepackage[breaklinks=true,unicode=true,urlcolor = blue,colorlinks = true,citecolor = blue,linkcolor = blue]{hyperref}
\usepackage{graphicx}
\usepackage{todonotes}
\usepackage{wrapfig}
\usepackage{easyReview}
\usepackage{cleveref}
\usepackage{tikz}
\usetikzlibrary{snakes}
\usepackage{nicefrac}

\def\slashchar#1{\setbox0=\hbox{$#1$}           
	\dimen0=\wd0                                 
	\setbox1=\hbox{/} \dimen1=\wd1               
	\ifdim\dimen0>\dimen1                        
	\rlap{\hbox to \dimen0{\hfil/\hfil}}      
	#1                                        
	\else                                        
	\rlap{\hbox to \dimen1{\hfil$#1$\hfil}}   
	/                                         
	\fi}                       %

\DeclareMathAlphabet\mathbfcal{OMS}{cmsy}{b}{n}

\def\epsilonb{{\mbox{\boldmath $\epsilon$}}}
\def\pib{{\mbox{\boldmath $\pi$}}}

\newcommand{\red}{\textcolor{red}}
\newcommand{\blue}{\textcolor{blue} }
\newcommand{\angstrom}{\textup{\AA}}
\newcommand{\newtext}[1]{\textcolor[rgb]{0.00,0.50,1.00}{#1}}
\renewcommand{\i}{\text{i}}
\newcommand{\tfd}{\ket{\text{TFD}}}
\renewcommand{\d}[2][]{\text{d}^{#1}#2}
\newcommand{\poi}[2]{\begin{Bmatrix}#1,#2\end{Bmatrix}}
\newcommand{\Cal}[1]{{\cal #1}}
\newcommand{\1}{\mathds{1}}

\newcommand{\myroundedbrackets}[1]{\left(#1\right)}
\newcommand{\JK}[1]{\textcolor{red}{#1}}
\newcommand{\SB}[1]{\textcolor{magenta}{#1}}
\def\avg#1{\left\langle#1\right\rangle}
\def\bra#1{\left\langle#1\right|}
\def\ket#1{\left|#1\right\rangle}
\def\braket#1#2{\left\langle #1|#2\right\rangle}
\def\ketbra#1#2{\left|#1\right\rangle\left\langle#2\right|}
\def\abs#1{\left|#1\right|}
\def\Re{{\rm Re}}
\def\Im{{\rm Im}}
\def\sgn{{\rm sgn}}
\def\mode{\text{ }{\rm mod}\text{ }}
\def\trace#1{{\rm Tr}\left[#1\right]}
\def\Det#1{{\rm Det}\left(#1\right)}
\def\te{\mathrm{e}}
\def\eff{\mathrm{eff}}
\def\tr{\mathrm{tr}}
\def\Tr{\mathrm{Tr}}
\def\TT{{T\bar T}}
\def\Id{\mathds{1}}
\def\Sch{{\text{Sch}}}
\def\tint{{\text{int}}}
\def\pq#1#2{\{#1,#2\}}
\def\TFD{\text{TFD}}

\def\nn{\nonumber}
\def\pa{\partial}
\DeclareMathOperator\arctanh{arctanh}
\DeclareMathOperator\arccosh{arccosh}
\DeclareMathOperator\arcsinh{arcsinh}
\DeclareMathOperator\sinc{sinc}

\begin{document}

	\title{Non-Locality induces Isometry and Factorisation in Holography}

    \author{Souvik Banerjee}
	
	\affiliation{Institute for Theoretical Physics and Astrophysics,
		Julius-Maximilians-Universit\"at W\"urzburg, 97074 W\"urzburg, Germany}
	
	\affiliation{W\"urzburg-Dresden Cluster of Excellence ct.qmat}
	
	\author{Johanna Erdmenger}
	
	\affiliation{Institute for Theoretical Physics and Astrophysics,
		Julius-Maximilians-Universit\"at W\"urzburg, 97074 W\"urzburg, Germany}
	
	\affiliation{W\"urzburg-Dresden Cluster of Excellence ct.qmat}
	
	\author{Jonathan Karl}
	
	\affiliation{Institute for Theoretical Physics and Astrophysics,
		Julius-Maximilians-Universit\"at W\"urzburg, 97074 W\"urzburg, Germany}
	
	\affiliation{W\"urzburg-Dresden Cluster of Excellence ct.qmat}

\begin{abstract}
\noindent

\noindent In holography, two manifestations of the black hole information paradox are given by the non-isometric nature of the bulk-boundary map and by the factorisation puzzle. By considering time-shifted microstates of the eternal black hole, we demonstrate that both these puzzles may be simultaneously resolved by taking into account non-local quantum corrections that correspond to wormholes arising from state averaging. This is achieved by showing, using a resolvent technique, that the resulting Hilbert space for an eternal black hole in Anti-de Sitter space is finite-dimensional with a discrete energy spectrum.  The latter gives rise to a transition to a type I von Neumann algebra. 
\end{abstract}

	\maketitle

\footnotetext[1]{Holographic CFTs are characterized by large central charges and sparse spectrum of operators with small conformal dimensions. Furthermore, correlators of these low-lying operators in any holographic CFT factorize into products of two-point functions, in other words connected higher point correlators of these operators vanish \cite{El-Showk:2011yvt}.}
\footnotetext[2]{We refer to the semiclassical description, together with perturbative corrections as the effective description, while the full quantum gravity theory is the fundamental description, including non-perturbative corrections. The fundamental description is also sometimes termed as the boundary description.}
\footnotetext[3]{In general the microstate of a thermodynamical system is defined as a pure state $\ket{\psi}$, which is indistinguishable from the equilibrium state represented by a thermal density matrix $\rho$, when probed by simple operators \cite{Climent:2024trz,Geng:2024jmm}, in the sense that
\begin{equation*}
    \bra{\psi}\mathcal{O}_1...\mathcal{O}_n\ket{\psi}=\Tr(\rho\,\mathcal{O}_1...\mathcal{O}_n)\,.
    \label{eq:microstate_condition}
\end{equation*}}

\footnotetext[4]{This method was also used to give a microscopic interpretation of Bekenstein-Hawking entropy by considering a different family of microstates \cite{Balasubramanian:2022gmo,Balasubramanian:2022lnw,Climent:2024trz,Geng:2024jmm}}

\footnotetext[5]{The rank of the Gram matrix may in principle directly be determined by either performing a Gram-Schmidt orthogonalization procedure, or equivalently by diagonalizing $G$. Since we have to find $d_\Omega$ in the limit $\Omega\to\infty$, this requires diagonalizing a matrix of infinite size, which is very hard. The resolvent method provides a simpler way to compute the eigenvalue density of the Gram matrix.}

\footnotetext[6]{A similar structure of the boundary Hilbert space was studied \cite{Leutheusser:2022bgi}, where general states with a gravity dual were considered.}

\footnotetext[7]{In the present example of an eternal black hole, the phases arising from the global time evolution can be measured if two observers in a gedanken experimental set-up perform a non-local measurement by jumping into the black hole from different sides. They would detect a time-shift while comparing their respective times at their meeting point in the bulk, provided their watches were initially synchronized at asymptotic infinity \cite{Verlinde:2020upt}.}

According to the holographic principle \cite{tHooft:1993dmi, Susskind:1994vu}, a class of quantum field theories, namely large $N$ holographic conformal field theories (CFT) \cite{Heemskerk:2009pn, El-Showk:2011yvt, Note1} in $d$ spacetime dimensions, possess dual descriptions in terms of semiclassical theories of gravity in asymptotically Anti-de Sitter (AdS) spacetimes in $d+1$ dimensions \cite{Maldacena:1997re, Witten:1998qj, Gubser:1998bc}. While this duality is extremely appealing for providing a geometric visualization of thermodynamics \cite{Witten:1998zw, Chamblin:1999hg, Chamblin:1999tk} and information processing \cite{Ryu:2006bv, Ryu:2006ef, Hubeny:2007xt} in semiclassical black holes, it leads to several paradoxes questioning the very possibility of a consistent quantum mechanical theory of gravity. These puzzles, referred to as information paradox in general, can be sharpened by using the quantitative equivalence between the two theories conjectured within the  duality. In particular, the identification of the partition functions between both theories appears to be inconsistent. 

In the present work we deal with two apparently distinct manifestations of this puzzle. The first one 
is the {\it non-isometric nature of the bulk-boundary map}. This is a statement of the fact that in the semiclassical limit characterized by perturbative quantum fields propagating in a fixed background, the bulk Hilbert space turns out to be much larger than its holographic counterpart \cite{Akers:2022qdl,Faulkner:2022ada, DeWolfe:2023iuq,Antonini:2024yif}. 

\begin{figure}
    \centering
    \includegraphics[width=1.0\linewidth]{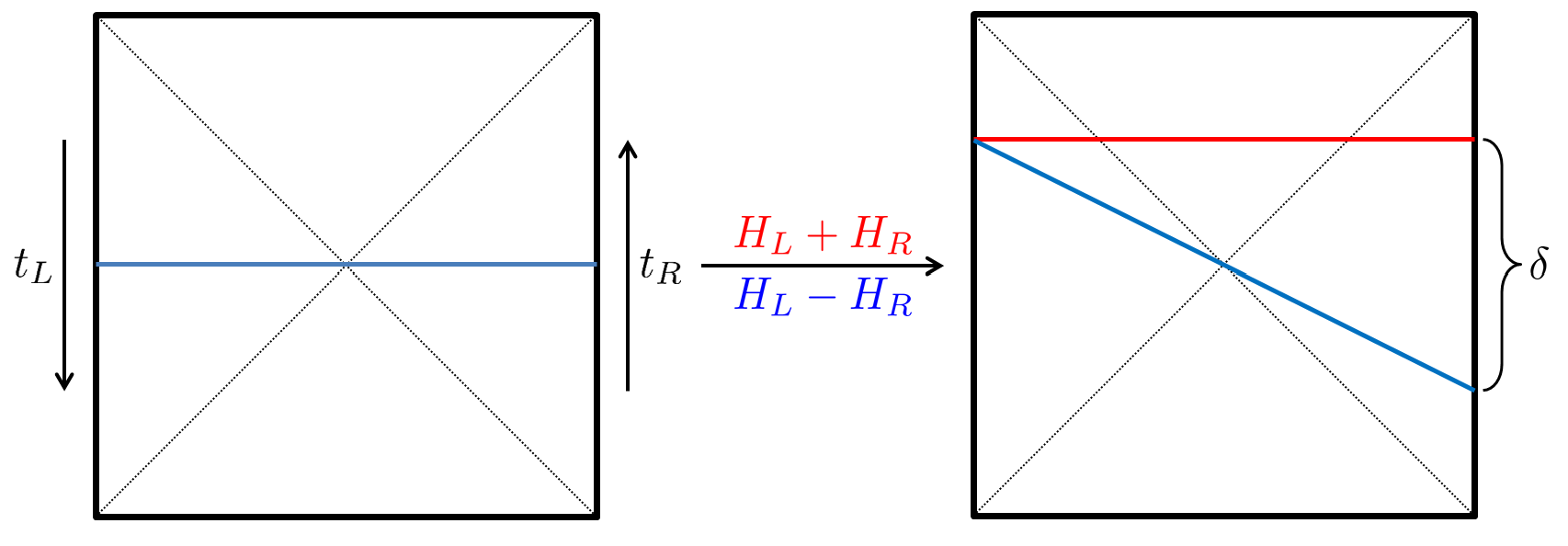} 
        \caption{Visualization of the black hole microstates \eqref{eq:generalised_TFD}. The left figure shows the Kruskal diagram of an eternal black hole. The vertical arrows indicate that time runs in different directions for the left and right boundary, while the black hole horizon is depicted by dashed lines. The TFD state \eqref{eq:TFD_state} corresponds to this geometry, together with an identification of the left and right boundary times along the Cauchy slice drawn in blue. This identification is preserved under the evolution with the difference of boundary Hamiltonians, shown on the right hand side. Evolving with the sum of the boundary Hamiltonians, as indicated by the red line, induces a time shift $\delta$, and the TFD state is changed to a generalized TFD state \eqref{eq:generalised_TFD}.} 
        \label{Fig:Eternal_BH}
\end{figure}

The second manifestation of the information paradox we focus on  is the {\it factorisation puzzle} \cite{Harlow:2018tqv, Jafferis:2019wkd}. This puzzle manifests itself in conflicting statements about the Hilbert space structures expected from the bulk and boundary perspectives of a two-sided eternal black hole in AdS. Following the ER = EPR proposal \cite{Maldacena:2013xja, VanRaamsdonk:2010pw}, the black hole is dual to two entangled CFTs living on the two boundaries \cite{Maldacena:2001kr} as depicted in Fig.~\ref{Fig:Eternal_BH}. Since the CFTs do not interact, the corresponding quantum state belongs to a factorized Hilbert space. The bulk Hilbert space on the other hand is not factorized due to the presence of geometric wormholes, resulting in an obvious mismatch between the partition functions of the bulk and the boundary theories. 
The factorisation of the Hilbert space may also be rephrased in terms of von Neumann (vN) algebras. The algebra of observables in the semiclassical bulk description of gravity is of type III$_1$ \cite{Book:Haag}. A type III algebra does not admit any irreducible representation, and therefore the Hilbert space on which it acts cannot be written as a tensor product. Nevertheless, it is expected that if it were possible to take into account all non-perturbative quantum corrections, the algebra of observables would eventually become type I, which admits irreducible representations \cite{Witten:2021unn}. Accordingly, a type I algebra allows pure black hole microstates which now belong to a factorized Hilbert space. This in principle solves the factorisation puzzle.

In this work we demonstrate that these two faces of the information paradox are actually connected and can be resolved together by invoking the fundamentally non-local nature of the quantum theory of gravity.
We achieve this goal by considering an infinite family of states,  
namely time-shifted thermofield double (TFD) states \cite{Papadodimas:2015xma, Papadodimas:2015jra, Verlinde:2020upt, Nogueira:2021ngh, Banerjee:2022jnv, Banerjee:2023liw, Banerjee:2023eew}, 
\begin{equation}
    \ket{\text{TFD}_\alpha}=\frac{1}{\sqrt{Z(\beta)}}\sum\limits_{n}e^{-\frac{\beta}{2}E_n}\,e^{i\alpha_n}\vert E_n\rangle_R\vert E_n\rangle_L\,,
    \label{eq:generalised_TFD}
\end{equation}
where the $\alpha_n$ are additional phases.
These states are entanglement-equivalent to the TFD state defined for $\alpha_n = 0$,
\begin{equation}
    \tfd=\frac{1}{\sqrt{Z(\beta)}}\sum\limits_ne^{-\frac{\beta}{2}E_n}\ket{E_n}_R\ket{E_n}_L\,,
    \label{eq:TFD_state}
\end{equation}
 where $Z(\beta) =  \sum_n e^{-\beta E_n}$ is the canonical thermal partition function of one boundary theory, and the $\vert E_n\rangle_{R/L}$ are the energy eigenstates of the right and left boundary theory, with corresponding energy eigenvalue $E_{n}$. The generalized TFD states \eqref{eq:generalised_TFD} are holographically dual to an eternal black hole in AdS, with different gluing of the left and the right boundaries to the bulk spacetime, as depicted in figure \ref{Fig:Eternal_BH} \cite{Papadodimas:2015xma, Papadodimas:2015jra}. These details about gluing of asymptotic boundaries cannot be detected by any bulk observer infalling from any of the boundaries and therefore quantify the inherent non-local nature of gravitational spacetimes. Moreover, this duality between generalised TFD states and the eternal black hole holds in any dimension.

To leading order in the large $N$ expansion, or equivalently in the expansion in Newton's constant $G_N$, these generalized TFD states are orthogonal to each other \cite{Verlinde:2020upt}. Consequently, they span a Hilbert space of infinite dimension. To leading order in $G_N$, this is a violation of the central dogma of black hole physics that the entropy contained in a space region is bounded by the surface area of that region, and  hence the  dimension of the black hole Hilbert space is finite. This mismatch of dimensionality between the bulk and boundary Hilbert spaces is referred to as non-isometric map.

We address this mismatch in the following way. Considering corrections non-perturbative in $G_N$, we show that the black hole microstates \eqref{eq:generalised_TFD}  have exponentially small overlaps. These can be interpreted as Euclidean replica wormhole corrections to the semiclassical gravity path integral. In particular, we show that these replica wormholes arise from state-averaging over the Hilbert space of phase-shifted TFD states. Since these states are dual to a Lorentzian two-sided black holes, this demonstrates a novel connection between two seemingly different notions of Euclidean and Lorentzian non-locality. By counting only the linearly independent microstates, we find that the dimension of the black hole Hilbert space is reduced to $D=e^{S_{\text{BH}}}$, where $S_{\text{BH}}$ is the Bekenstein-Hawking entropy of the black hole, in agreement with the central dogma. 
{This demonstrates how the isometric nature of the bulk-boundary map is retrieved through state-averaged replica wormholes.} 
The central point here is that the reduction in the dimension of the Hilbert space is achieved via wormhole contributions that are manifestly non-local in the Lorentzian ER=EPR sense. Furthermore they explicitly take into account the state dependence in the holographic bulk reconstruction \cite{Papadodimas:2015xma, Papadodimas:2015jra}. Interestingly, this reduction in the dimension of Hilbert space through state dependence is consistent with the proposal made in \cite{Antonini:2024yif} that  relates state-dependence to non-isometric maps. 
Moreover, our argument does not require the inclusion of additional external degrees of freedom such as branes \cite{Penington:2019kki,Brown:2019rox,Geng:2024jmm} or matter fields \cite{Balasubramanian:2022gmo,Balasubramanian:2022lnw,Climent:2024trz}. Instead we  only rely on the fundamental non-locality of gravity.

Furthermore, we show how the aforementioned reduction in the dimension of the Hilbert space can serve as a simple demonstration of 
how the type III$_1$  vN algebra associated to the bulk undergoes a transition to a type I$_D$ vN algebra: the latter is given by all operators acting on the black hole Hilbert space of dimension $D=e^{S_{\text{BH}}}$ once the non-perturbative corrections are included. We show that 
the transition to type I$_D$ is  a consequence of the energy spectrum becoming discrete. The countable energy eigenstates then span a factorized Hilbert space. {This demonstrates how the factorisation puzzle can be resolved by the non-local nature of quantum gravity}. In particular, we show directly at the level of the Hilbert space how factorisation is achieved, without relying on the factorisation of the two-point function \cite{Boruch:2024kvv,Balasubramanian:2024yxk,Li:2024nft}.

While we were preparing for submission, the work \cite{Magan:2024nkr} appeared on arXiv which also discusses the resolvent computation in the context of time-shifted TFD states.

\section{Brief review of generalized TFD states}
\label{sec:microstates}

\noindent From the perspective of the boundary theory which is an ordinary quantum theory without gravity, the TFD state \eqref{eq:TFD_state} can be thought of as a split state \cite{Roos:1970fm,Buchholz:1973bk,fewster2016split} satisfying
\begin{equation}
    \Tr(\rho_{\rm TFD} A_LB_R)=\Tr_L(\rho_L A_L)\,\Tr_R(\rho_R B_R)\,,
    \label{eq:split property}
\end{equation}
where $\rho_{\rm TFD} = \tfd\langle {\text {TFD}}|$ is the TFD density matrix. $\rho_{L/R}$ are density matrices, and $A_L, B_R$ are observables defined for the left and right system respectively. In this way the TFD state \eqref{eq:TFD_state} can be thought of as the purification of a thermal state represented by the thermal density matrix $\rho_R$. From this perspective, the von Neumann entropy associated with $\rho_R$ may also be interpreted as the entanglement entropy between the left and the right systems. The interpretation of the TFD state as a split state is a more precise restatement of the factorisation of the boundary Hilbert space.

The ER=EPR proposal \cite{Maldacena:2013xja, VanRaamsdonk:2010pw} provides a holographic bulk interpretation of the TFD state as the Hartle-Hawking wave function of a two-sided black hole, i.e.~a wormhole in AdS \cite{Maldacena:2001kr}. Consequently, the 
entanglement entropy associated with the reduced density matrix $\rho_R$ is identified with the thermal entropy of the two-sided eternal black hole \cite{Verlinde:2020upt}
\begin{equation}
    S(\rho_{R})=-\Tr_R\,\rho_R\log(\rho_R)=S_{\text{BH}}=\frac{A}{4G_N}\,,
    \label{eq:BH_entropy}
\end{equation}
where $A$ is the area of the black hole horizon. This implies that the thermal partition function $Z(\beta)$ appearing in \eqref{eq:TFD_state} corresponds to the Gibbons-Hawking partition function \cite{PhysRevD.15.2752}. We note that this interpretation crucially relies on the fact that for any local quantum mechanical system, such as the boundary CFT, the split property is satisfied, and consequently reduced density matrices exist.

However, the gravity theory in the bulk is manifestly non-local. Therefore the split property, which is a statement about locality, ceases to exist in the bulk \cite{Raju:2021lwh}. In particular, this implies that density matrices associated with a local subregion do not exist in gravity. Consequently, the TFD state dual to such a wormhole geometry can no longer be identified as the purification of a thermal state, and the identification \eqref{eq:BH_entropy} is does not hold, since the left hand side of that equation is not well defined.

The non-locality of gravity manifests itself through the absence of a time-reversal symmetry, or equivalently an origin of time. For the two-sided eternal black hole in AdS, this non-local nature of gravity results in a large class of states, namely the generalised TFD states \eqref{eq:generalised_TFD}. These states arise from one-sided time evolutions of the TFD state 
\begin{equation}
    \ket{\text{TFD}_\alpha}=e^{iH_Lt}\vert \text{TFD}\rangle\,,
    \label{eq:evolved_TFD}
\end{equation}
identifying $\alpha_n=E_nt$, $H_L$ being the left boundary Hamiltonian. These states share the  same dual black hole geometry, nevertheless, with different identifications of the left and the right boundary points. The original TFD state \eqref{eq:TFD_state} corresponds to one particular identification of boundary times, which is along the Cauchy slice shown in blue on the left-hand side of Fig.~\ref{Fig:Eternal_BH}, with  support at $t_L = t_R = 0$.

For observers infalling from one of the boundaries, all the states \eqref{eq:evolved_TFD} yield the same correlation functions \cite{Papadodimas:2015xma, Papadodimas:2015jra}. This fact reflects, quantitatively, the freedom to  independently choose the origins of time for the left and right boundary theory.  Since time runs in different directions in the left and right exterior, this identification is respected by time evolution generated by the difference of the left and right boundary Hamiltonians
\begin{equation}
    U_-(t)=e^{i\frac{t}{2}(H_L- H_R)}\,,
    \label{eq:evolution_difference}
\end{equation}
in the sense that the TFD state with a given phase remains invariant under this evolution. A direct consequence of this fact is that we can also obtain the generalised TFD states \eqref{eq:evolved_TFD} from global time evolution, which is generated by 
\begin{equation}
    \ket{\text{TFD}_\alpha}=U_+(t)\, U_-(t)\vert \text{TFD}\rangle=U_+(t)\vert \text{TFD}\rangle\,,
\end{equation}
where the unitary transformation $U_+$ is given by the sum of the boundary Hamiltonians,
\begin{equation}
    U_+(t)=e^{i\frac{t}{2}(H_L+ H_R)}\,.
    \label{eq:unitaries}
\end{equation}
The global time evolution is illustrated on the right-hand side of Fig.~\eqref{Fig:Eternal_BH}. Since the evolved Cauchy slice contains a larger portion of the black hole interior, we may also interpret the evolved state as being dual to a Lorentzian wormhole of different length. 

Every state in the infinite family \eqref{eq:generalised_TFD} is a microstate of the black hole \cite{Note3}, which shows that the bulk Hilbert space appears to have way too many states when probed by a local observer. However, it is worth emphasizing that although the relative time-shift between the left and the right boundaries remains undetected by a single infalling observer, the phases can be detected through non-local measurements \cite{Verlinde:2020upt,Note7}. In what follows, we will demonstrate how non-local effects can actually lift this infinite microstate degeneracy.

\section{Microstates and overlaps}
\noindent We now show that the infinitely many black hole microstates \eqref{eq:generalised_TFD} have exponentially small overlaps, which arise from wormholes, i.e.~non-local contributions in the gravitational path integral. 

In our case, the non-local nature of gravity is encoded in the phases of the generalised TFD states \eqref{eq:generalised_TFD}. At large $N$, and large time of the order of the Poincaré recurrence time \cite{Verlinde:2020upt}, these phases are random, since they are related to the random energy spectrum of the holographic CFT \eqref{eq:evolved_TFD}. 
Therefore, the states satisfy an orthogonality condition of the form
\begin{equation} \langle\text{TFD}_\alpha\vert\text{TFD}_\gamma\rangle=\sum\limits_np^2_ne^{i(\gamma_n-\alpha_n)}=\delta_{\alpha\gamma}+\mathcal{O}(e^{-S_{\text{BH}}/2})\,,
    \label{eq:orthogonal}
\end{equation}where  $p_n=e^{-\beta E_n/2}/\sqrt{Z}$ is the Boltzmann probability \cite{Verlinde:2020upt}. This equation implies that in the semiclassical limit, these phase-shifted microstates form an infinite basis of the black hole Hilbert space at late enough times. 

In Euclidean gravity the phase-shifted TFD state corresponds to the Hartle-Hawking state representing an Euclidean eternal black hole with a given identification of the asymptotic boundaries \cite{Maldacena:2001kr}. The overlap \eqref{eq:orthogonal} between these states appears as the leading order contribution in the topological expansion of the Euclidean gravitational path integral (see supplemental material for more details). The different identifications between boundary times manifest through an ambiguity in imposing boundary field data on this path integral. This ambiguity is associated with the lack of a manifest time reversal symmetry, or equivalently, the bulk non-locality giving rise to the additional phases \eqref{eq:evolved_TFD}. Due to this non-locality the orthogonality relation \eqref{eq:orthogonal} receives exponentially corrections. In the topological expansion of the gravitational path integral these quantum corrections appear as connected geometries. To show this, we compute the second moment
 \begin{align}
    \label{eq:derivation_overlaps}
    &\frac{1}{\mathcal N}\sum\limits_\gamma\vert\langle\text{TFD}_{\alpha}\vert\text{TFD}_\gamma\rangle\vert^2
        =\frac{\delta_{\alpha\alpha}}{\mathcal N} + \sum\limits_{n}p_n^4= \frac{1}{\mathcal N}+ \frac{Z(2\beta)}{Z^2(\beta)}\,,
    \end{align}

\noindent where $p_n=e^{-\beta E_n/2}/\sqrt{Z}$ is the Boltzmann probability, with $Z(\beta)$  the thermal partition function, and $\mathcal{N}$ is the number of phases contributing to the sum. In the derivation of \eqref{eq:derivation_overlaps} we have explicitly used the randomness of the phases at large time scale. The details of this calculation are included in the supplemental material.

\begin{figure}
    \includegraphics[width=1.0\linewidth]{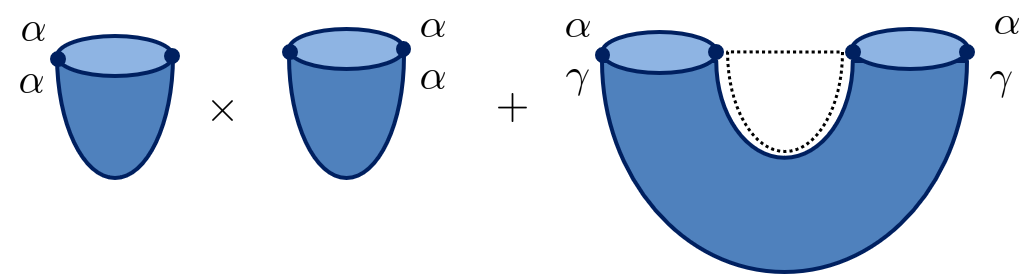}.
      \caption{Geometric representation of the right hand side of \eqref{eq:derivation_overlaps}. The first term corresponds to two disconnected topologies, while the second represents a replica wormhole arising from state averaging. The averaging over the states labeled by $\gamma$ is depicted through the dashed line in this diagram.}
    \label{eq:wormhole}
\end{figure}

The first term in \eqref{eq:derivation_overlaps} arises from the square of \eqref{eq:orthogonal}, which from the point of the gravitational path integral corresponds to the disconnected contribution, given by two copies of an Euclidean black hole. The second term in \eqref{eq:derivation_overlaps}, given by the ratio of partition functions, arises from a Euclidean wormhole connecting two boundaries, known as replica wormhole \cite{Penington:2019kki,Verlinde:2020upt}. In particular, the second term in \eqref{eq:derivation_overlaps} represents the properly normalized partition function of this wormhole \cite{Verlinde:2021kgt}. Schematically, the topological expansion corresponding to \eqref{eq:derivation_overlaps} is represented as shown in Fig.~\ref{eq:wormhole}.
These wormhole corrections to the overlap are non-perturbative in nature, since they are exponential in the black hole entropy, that is of ${\cal O}\left(e^{-S_{\rm{BH}}}\right)$. Equivalently, they are non-perturbative in an expansion in Newton's constant. \\
The appearance of  connected contributions in \eqref{eq:derivation_overlaps} may also be explained as arising from an averaging procedure. From that perspective, the quantum amplitude  $\langle\text{TFD}_\alpha\vert\text{TFD}_\gamma\rangle$ is interpreted as the expectation value of a random matrix
\begin{equation}
    {\mathcal M}_{\alpha\gamma}=\delta_{\alpha\gamma}+e^{-S_{\text{BH}}/2}\,\mathcal{R}_{\alpha\gamma}\,,
    \label{eq:random matrix}
\end{equation}
where $\mathcal{R}_{\alpha\gamma}$ is a random matrix with zero mean \cite{Penington:2019kki}. This reproduces \eqref{eq:orthogonal} when taking the average of this matrix $\mathcal{M}$. In its matrix quantum mechanical description \eqref{eq:random matrix}, the higher moment \eqref{eq:derivation_overlaps} represents the variance of $\mathcal{R}$. Thus, the gravitational path integral computes an average over the fundamental degrees of freedom \cite{Note2} that are encoded in the random matrix \cite{Penington:2019kki}. In particular, in our case the averaging over $\mathcal{M}$ is a state averaging over the Hibert space of phase-shifted states.

Note that the sum in \eqref{eq:derivation_overlaps} is taken over the phases appearing in \eqref{eq:generalised_TFD}, which have a Lorentzian interpretation through \eqref{eq:evolved_TFD}. Each of the phase-shifted states \eqref{eq:generalised_TFD} then has a dual gravitational description in terms of a Lorentzian wormhole of a given length. These phases are due to bulk non-locality as they are not detectable by any local measurements, and thus manifest the non-local features associated with Lorentzian Einstein-Rosen wormholes. However in order to derive \eqref{eq:derivation_overlaps}, we only need the definition \eqref{eq:generalised_TFD} without any reference to the Lorentzian origin of these phases. This brings us to a salient feature of our construction which prompts that, even though the phases appearing in \eqref{eq:generalised_TFD} manifest Lorentzian non-locality, at the level of gravitational path integral, the same phases contribute to an Euclidean notion of non-locality through the replica wormhole contributions given by \eqref{eq:derivation_overlaps}. These two notions of Lorentzian and Euclidean wormholes, and hence the associated interpretations of non-locality are a priori different and their relation had so far been elusive. {Our choice of microstates, and their interpretation given above, thus demonstrates how replica wormholes arise from an average over states corresponding to different Lorentzian wormholes, thereby connecting the roles played by Lorentzian and Euclidean wormholes in manifesting non-locality in a quantum theory of gravity.}

\section{Counting the microstates}
\label{sec:Microstate_counting}

\noindent Due to \eqref{eq:derivation_overlaps}, not all of the infinitely many microstates \eqref{eq:generalised_TFD} contribute to the dimension of the black hole Hilbert space $\mathcal{H}_\text{BH}$. We now show that the true dimension of this Hilbert space  is finite and given by $D=e^{S_{\text{BH}}}$. 

For this purpose we count the number of linearly independent microstates in a fixed microcanonical energy window $[E_0,E_0+\Delta E]$, where the microstates take the form \cite{Chandrasekaran:2022eqq}
\begin{equation}
    \ket{\psi_i}=\sum\limits_{n}e^{-\beta(E_n-E_0)/2}\,e^{i\alpha_{i,n}}f(E_n-E_0)\ket{E_n}_L\ket{E_n}_R\,.
    \label{eq:microcanonical_TFD}
\end{equation}
Here $f$ is an arbitrary smooth function, supported in the respective energy window, and $f$ is chosen such that the states \eqref{eq:microcanonical_TFD} are normalized. In the large $N$ limit the microcanonical TFD states are indistinguishable from their canonical counterpart, since correlation functions for simple operators in these states match \cite{Chandrasekaran:2022eqq}.
In order to count the microstates we consider the Hilbert space
\begin{equation}
     \mathcal{H}_\Omega=\text{span}\{\,\vert\psi_i\rangle\,\vert\,i=1,...,\Omega\,\}\,,
     \label{eq:subset}
\end{equation}
spanned by $\Omega$ of the black hole microstates, and analyse its dimension $d_\Omega$ as a function of $\Omega$. Since $\mathcal{H}_\Omega$ is a subspace of $\mathcal{H}_\text{BH}$ we have
\begin{equation}
    d_\Omega:=\text{dim}\,(\mathcal{H}_\Omega)\leq\text{min}(D,\Omega)\,,
\end{equation}
and the dimension of the black hole Hilbert space is obtained from the limit value of $d_\Omega$ as $\Omega\to\infty$. We determine $d_\Omega$ by counting the number of linearly independent generalised TFD states in $\mathcal{H}_\Omega$. For this purpose, we introduce the Gram matrix $G$ with entries
\begin{equation}
    G_{ij}=\langle\psi_i\vert\psi_j\rangle\,,\quad i,j=1,...,\Omega\,,
    \label{eq:Gram_matrix}
\end{equation}
which is positive semi-definite by construction.  The rank, i.e.~the number of positive eigenvalues of the Gram matrix, is then equal to $d_\Omega$ \cite{Note5}. Since the Gram matrix may be interpreted as a random matrix \eqref{eq:random matrix}, as explained in the previous section, a suitable method to determine $d_\Omega$ is the resolvent method \cite{Note4,CVITANOVIC198149,2009arXiv0911.0087S,Penington:2019kki}. We use this method in order to find the eigenvalue density of $G$, which is defined as
\begin{equation}
    D(\lambda):=\text{Tr}\,(\delta(\lambda\Id-G))=\sum\limits_{i=1}^\Omega\,\delta(\lambda-\lambda_i)\,,
    \label{eq:eigenvalue_density}
\end{equation}
where the $\lambda_i$ are the eigenvalues of $G$. The rank of the Gram matrix is then obtained from the eigenvalue density as
\begin{equation}
    d_\Omega=\lim\limits_{\epsilon\to0^+}\int\limits_{\epsilon}^\infty d\lambda\,D(\lambda)=\Omega-\text{Ker}(G)\,,
    \label{eq:rank_Gram_matrix}
\end{equation}
where Ker$(G)$ is the dimension of the kernel of the Gram matrix, and the second equality is the rank-nullity theorem. The dimension of the kernel corresponds to the number of zero eigenvalues of $G$, which can be obtained from \eqref{eq:eigenvalue_density}. Using the resolvent method \cite{CVITANOVIC198149,2009arXiv0911.0087S,Penington:2019kki}, we calculate
the explicit form of the eigenvalue density and find
\begin{align}
    \label{eq:result_EV density}
        &D(\lambda)=\delta(\lambda)\left(\Omega-e^{S_{\text{BH}}}\right)\Theta\left(\Omega-e^{S_{\text{BH}}}\right)+\nonumber\\
        &\frac{e^{S_{\text{BH}}}}{2\pi\lambda}\sqrt{\left[\lambda-\left(1-\sqrt{\frac{\Omega}{ e^{S_{\text{BH}}}}}\right)^2\right]\left[\left(1+\sqrt{\frac{\Omega}{ e^{S_{\text{BH}}}}}\right)^2-\lambda\right]}\,.\nonumber\\
\end{align}
The details of this calculation are given in the supplemental material. The singular term of the eigenvalue density \eqref{eq:result_EV density}, i.e.~the term proportional to $\delta(\lambda)$, corresponds to the zero eigenvalues of $G$, while the continuous term gives the distribution of the positive eigenvalues. We note that the appearance of the Bekenstein-Hawking entropy \eqref{eq:BH_entropy} in \eqref{eq:result_EV density} arises from the overlaps \eqref{eq:derivation_overlaps}, since $S_\text{BH}$ is contained in the ratio of the Gibbons-Hawking partition functions.  These overlaps are encoded in the higher moments of the Gram matrix that appear in the definition of the resolvent \eqref{eq:resolvent_matrix}. 

Combining \eqref{eq:result_EV density}, with \eqref{eq:rank_Gram_matrix}, we finally determine the rank of $G$, i.e.~the dimension of the Hilbert space $\mathcal{H}_\Omega$ as 
\begin{equation}
    \begin{split}
    d_\Omega&=\min\left(\Omega,e^{S_{\text{BH}}}\right)\,.
    \end{split}
    \label{eq:dimension_Hilbert_space}
\end{equation}
The result of this calculation shows that if the number of microstates $\Omega$ in the Hilbert space $\mathcal{H}_\Omega$ is smaller than $e^{S_{\text{BH}}}$, the microstates span a proper subspace of the black hole Hilbert space of dimension $\Omega$. Moreover, following \eqref{eq:rank_Gram_matrix}, for this case the Gram matrix does not have zero eigenvalues, i.e.~it has a trivial kernel. Moreover, we find from \eqref{eq:dimension_Hilbert_space} for $\Omega>e^{S_{\text{BH}}}$ that the rank of the Gram matrix reaches a saturation value, given by the Bekenstein-Hawking entropy $e^{S_{\text{BH}}}$. This indeed shows that the black hole Hilbert space has dimension $D=e^{S_{\text{BH}}}$. 

\section{Non-isometric map}
\noindent The results of the previous section show in an explicit example how the Hilbert space of the effective theory $\mathcal{H}_\text{eff}$ can be reduced to the Hilbert space of the fundamental theory $\mathcal{H}_{\text{BH}}$ \cite{Note2} upon inclusion of non-local features of quantum gravity. In the effective description, \eqref{eq:orthogonal} does not receive  corrections, and thus the Hilbert space is spanned by all of the infinitely many black hole microstates, while the fundamental Hilbert space is spanned by finitely many microstates \eqref{eq:dimension_Hilbert_space}.
This shows explicitly that the embedding map is non-isometric \cite{Akers:2022qdl,Faulkner:2022ada}
\begin{align}
\label{eq:nonisometricmap}
{\cal F}_{\rm embedding} : {\cal H}_\text{eff} \rightarrow {\cal H}_{\rm {BH}}\,.
\end{align}
The non-isometric nature of the map implies the existence of a large class of null states that it annihilates. Following the derivation in the previous section, this set of null states is straightforwardly identified with the kernel of the Gram matrix \eqref{eq:Gram_matrix}. In particular, it follows directly from \eqref{eq:result_EV density} that this kernel is non-trivial, as soon as $\Omega=e^{S_{\text{BH}}}$ states are included in $\mathcal{H}_\Omega$. In order to make sense of the low energy effective field theory it is necessary that the existence of this null subspace cannot be detected by a local observer, having access to simple measurements, but instead requires an operation of exponential complexity. In our case, this directly follows from the non-locality of the connected contributions in state overlaps \eqref{eq:derivation_overlaps},  which are indeed of ${\cal O}\left(e^{-S_{\rm{BH}}}\right)$. It is worth emphasizing that in our approach, the non-isometric map does not require the introduction of additional external non-gravitational degrees of freedom. It naturally follows through exploiting inherent non-locality of quantum gravity.

{To leading order in $G_N$, the number of independent microstates in the effective Hilbert space violates the holographic entropy bound of \cite{Bousso:2002ju}. 
The non-isometric map considered here shows that this
direct consequence of the emergent bulk locality in the low-energy description. Including the quantum corrections ensures that the bound is satisfied.}

\section{Algebra and Factorisation}
\label{sec:vN_algebras}
We now provide an algebraic interpretation of the aforementioned non-isometric map, in particular, the algebraic definition of the effective Hilbert space and the fundamental (black hole) Hilbert space \cite{Note2}. In the process we show how these interpretations automatically resolve the factorisation puzzle.

The vN algebra of observables associated to semiclassical gravity is of type III$_1$, which is a consequence of the fact that in the $G_N\to0$ limit, or equivalently large $N$ limit, the energy spectrum becomes continuous since energy gaps scale as $1/N$ \cite{Furuya:2023fei}. Due to this continuity, or equivalently  due to the randomness of the energy spectrum, the microstates \eqref{eq:generalised_TFD} are orthogonal \eqref{eq:orthogonal}, and span an infinite-dimensional Hilbert space. In this limit the TFD state is no longer a split state \eqref{eq:split property}, and accordingly does not belong to a factorized Hilbert space $\mathcal{H}_L\otimes\mathcal{H}_R$. Instead the TFD state here corresponds to the vacuum state of a GNS Hilbert space \cite{Witten:2021jzq} given by
\begin{equation}
    \mathcal{H}_{\text{GNS}}=\{\,\ket{a}\,,a\in\mathcal{A}_L\,\vert\,\langle a\vert b\rangle:=\langle\text{TFD}\vert\, a^*\, b\tfd\,\}\, ,
    \label{eq:GNS_HS}
\end{equation}
where $\mathcal{A}_L$ is the type III$_1$ von Neumann algebra generated by the single trace operators of the left CFT \cite{Leutheusser:2021qhd,Leutheusser:2021frk}. The TFD state is associated to the identity operator of this type III$_1$ algebra, which does not admit irreducible representations. Consequently the GNS Hilbert space does not allow for pure states. Moreover a type III algebra prohibits defining a trace, and accordingly density matrices and von Neumann entropies cannot be defined. This is consistent with the fact that in the semiclassical limit, the TFD state cannot be identified with the purification of a thermal state.

\begin{figure}
    \centering
    \includegraphics[width=0.7\linewidth]{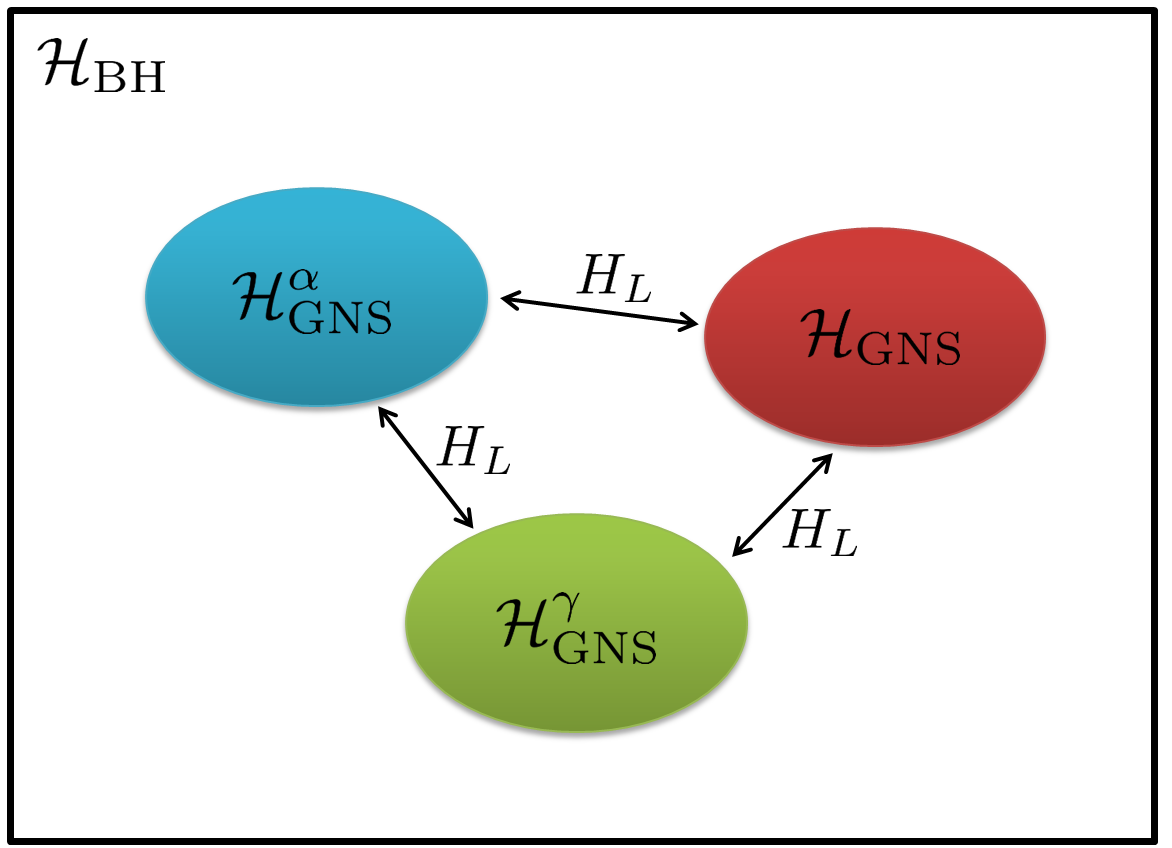} 
        \caption{At large $N$, the black-hole Hilbert space $\mathcal{H}_{\text{BH}}$ splits into GNS Hilbert spaces around the TFD state, labelled by $\mathcal{H}_{\text{GNS}}$, and the microstates \eqref{eq:generalised_TFD}, labelled $\mathcal{H}^{\alpha/\gamma}_{\text{GNS}}$. Due to \eqref{eq:evolved_TFD}, the left boundary Hamiltonian allows us to transition between different GNS Hilbert spaces.
        } 
        \label{Fig:BH_HS}
\end{figure}

It is worth emphasizing that each black hole microstates \eqref{eq:generalised_TFD} can be considered as the ground state of a different infinite dimensional GNS Hilbert space, associated with a fixed value of $\alpha$, defined as
\begin{equation}
    \mathcal{H}_{\text{GNS}}^\alpha=\{\,\ket{a}\,,a\in\mathcal{A}_L\,\vert\,\langle a\vert b\rangle:=\langle\text{TFD}_\alpha\vert\, a^*\, b\ket{\text{TFD}_\alpha}\,\}\,.
    \label{eq:alpha_GNS}
\end{equation}
Each of these Hilbert spaces with  fixed $\alpha$  can be thought of as a low-energy effective Hilbert space, corresponding to the Fock space of quantum fields propagating in curved spacetime. In this Hilbert space \eqref{eq:alpha_GNS} any local measurement is insensitive to $\alpha$. However, in order to transition between TFD states with different phases we need the Hamiltonian $H_L$ \eqref{eq:evolved_TFD}, which is not included in the type III algebra $\mathcal{A}_L$. Therefore, at large $N$ the black hole Hilbert space, which contains all TFD$_\alpha$ states, is given by a disjoint union of these different GNS Hilbert spaces (see Fig.~\ref{Fig:BH_HS}) \cite{Note6}. Since we work in the microcanonical ensemble \eqref{eq:microcanonical_TFD}, we can include the left boundary Hamiltonian $H_L$ into the algebra, which leads to a transition to a type II$_\infty$ algebra \cite{Chandrasekaran:2022eqq}. The extended algebra admits density matrices and renormalized von Neumann entropies, but pure states still do not exist. This algebra acts on an extended Hilbert space
\begin{equation}
    \hat{\mathcal{H}}=\mathcal{H}_{\text{GNS}}\otimes L^2(\mathbb{R})\,,
    \label{eq:type_II_HS}
\end{equation}
where $L^2$ are the square integrable functions. The resulting perturbative Hilbert space $\hat{\mathcal{H}}$ includes all black hole microstates \eqref{eq:generalised_TFD}, and corresponds to the infinite dimensional effective Hilbert space $\mathcal{H}_\text{eff}$ \eqref{eq:nonisometricmap}. 

We showed above that non-perturbative corrections break the orthogonality relation \eqref{eq:derivation_overlaps}, and consequently the black hole Hilbert space is finite dimensional \eqref{eq:dimension_Hilbert_space}. The algebra of observables associated to this Hilbert space is of type I$_D$, where $D=e^{S_\text{BH}}$. There is a simple yet precise way to justify this transition to a type I algebra through the discreteness of the energy spectrum \cite{Furuya:2023fei}. In our case the discreteness arises from wormhole corrections as follows. If we assume that the spectrum is continuous, the phases $\alpha_n =E_nt$ are random, and consequently as we argued in \eqref{eq:orthogonal} the microstates are orthogonal, leading to a type II algebra. By including non-perturbative wormhole corrections we find that the microstates \eqref{eq:generalised_TFD} are not orthogonal, and consequently the assumption that the spectrum is continuous has to break down. Thus, the energy spectrum is discrete, and we can actually write the TFD state in terms of direct products over energy eigenstates $\vert E_n\rangle_{L/R}$, which form the basis of a factorized Hilbert space 
    \begin{align}
        \mathcal{H}_L\otimes\mathcal{H}_R=\text{span}(\,\vert E_n\rangle_L\otimes \vert E_m\rangle_R\,)\,.
        \label{eq:factorized_HS}
    \end{align}
{This confirms in an explicit example the conjecture of \cite{Witten:2021unn} that a non-perturbative description of black holes leads to a transition to a type I algebra, and thereby resolves the factorisation puzzle.}

\begin{acknowledgments}
	{\it Acknowledgments ---}
	We would like to thank Arnab Kundu for collaboration during  early stages of this work, as well as Andreas Blommaert, Aidan Herderschee, Hao Geng, Sergio Hernandez-Cuenca, Thomas Kögel, Jonah Kudler-Flam, Dominik Neuenfeld, Suvrat Raju, Martin Sasieta, Ronak M.~Soni and Herman L.~Verlinde for useful discussions. We are grateful to Aidan Herderschee, Jonah Kudler-Flam, Suvrat Raju and Ronak M.~Soni for comments on an earlier version of the manuscript. S.B., J. E.~and J.K.~are supported by the Deutsche Forschungsgemeinschaft (DFG, German Research Foundation) through the German-Israeli Project Cooperation (DIP) grant ‘Holography and the Swampland’, as well as under Germany’s Excellence Strategy through the Würzburg-Dresden Cluster of Excellence on Complexity and Topology in Quantum Matter- ct.qmat (EXC 2147, project-id 390858490). J.E.~is grateful to the Massachusetts Institute of Technology, Princeton University and IAS Princeton for hospitality during the final stages of this work.
\end{acknowledgments}

\vspace{5mm}

\appendix

\section{\large{Supplemental material}}

\section{Path integral derivation of the TFD state}
\label{apx:PI}

Here we explain how the black hole microstates \eqref{eq:generalised_TFD}, as well as their overlaps \eqref{eq:orthogonal} and \eqref{eq:derivation_overlaps} can be obtained from the gravitational path integral. Therefore, we first calculate the Hartle-Hawking state of the two-sided eternal black hole \cite{PhysRevD.13.2188}, which is the holographic dual of the generalized TFD state. The Hartle-Hawking wave function is given by $\Psi_{\text{HH}}=\braket{\phi}{\Omega}$, where $\ket{\Omega}$ is the ground state at infinite past \cite{Harlow:2014yka}. This wave function is computed, up to a normalisation factor, by a Euclidean gravitational path integral
\begin{equation}
    \Psi_{\text{HH}}\sim\int\mathcal{D}g\mathcal{D}\phi\,e^{-S_E[g,\phi]}\,,
    \label{eq:Hartle-Hawking_state}
\end{equation}
where $\phi$ denotes a collection of matter field, which propagate in a geometry that is specified by the metric $g$. Furthermore $S_E$ is the Euclidean version of the Einstein-Hilbert action. We work in the semiclassical limit $G_N\to 0$ for which the path integral reduces to a sum over all classical solutions, which are consistent with the imposed boundary conditions, obtained from the dual CFT. 

\begin{figure}
    \centering
    \includegraphics[width=1.0\linewidth]{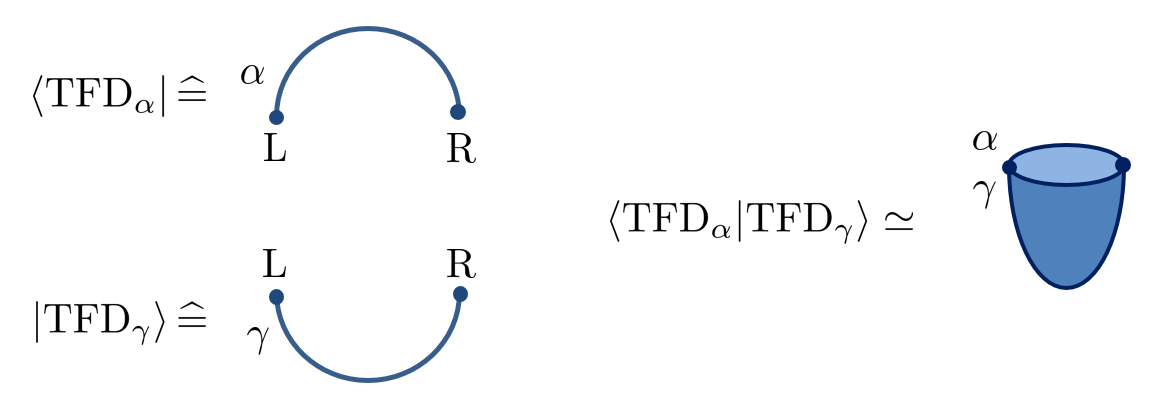} 
        \caption{The left-hand side shows the boundary conditions imposed in the gravitational path integral when preparing the bra or the ket of a generalized TFD state \eqref{eq:generalised_TFD}. These correspond to the upper or lower half of a thermal circle that is cut open along the $\tau=0$ point in Euclidean time. The phases enter the boundary conditions via the respective identification between left and right time. The boundary condition imposed on the path integral, when computing the overlap between different microstates \eqref{eq:orthogonal} corresponds to a single thermal circle, obtained by gluing both halves of the circle together. The leading geometry contributing to the gravitational path integral, which is consistent with this boundary condition is given by the Euclidean black hole, which is shown on the right hand side. Since different phases $\alpha/\gamma$ correspond to different identifications of boundary times this path integral gives zero, unless the phases coincide.} 
        \label{Fig:PI_Black_hole}
\end{figure}

The TFD state can be thought of as the square root of a thermal density matrix at temperature $\beta$ and is therefore prepared by a path integral over half a thermal circle, where we compactify Euclidean time \cite{Witten:2021jzq}. The spatial part of the boundary is given by $S^{d-1}$. Therefore on the CFT side we integrate over the manifold $I_{\beta/2}\times S^{d-1}$, where $I_{\beta/2}$ is an interval of length $\beta/2$ (see Fig.~\ref{Fig:PI_Black_hole}). From the gravity side the dominant saddle-point geometry which satisfies this boundary conditions is given by the lower half of the Euclidean Schwarzschild black hole in AdS, whose metric we denote by $\hat{g}$. This geometry may be thought of as the Euclidean black hole (see Fig.~\ref{Fig:PI_Black_hole}), which is cutoff at the $\tau=0$ surface. From this surface, the geometry can be analytically continued to yield the Lorentzian eternal AdS black hole, and the $\tau=0$ surface corresponds to a Cauchy slice, which passes through the bifurcation point (see Fig.~\ref{Fig:Eternal_BH}). The two event horizons emerge from the $r=0$ point of this Cauchy slice, where $r$ is the radial AdS coordinate. The ground state may then be obtained from a generic state $\ket{\psi}$ as
\begin{equation}
    \braket{\phi}{\Omega}=\frac{1}{\braket{\Omega}{\psi}}\lim\limits_{\tau\to\tau_I}\bra{\phi}e^{-\tau H}\ket{\psi}\,,
    \label{eq:Infinite_past}
\end{equation}
where $\tau_I$ is an initial value of Euclidean time, and up to a normalisation factor \eqref{eq:Infinite_past} is computed by the path integral
\begin{equation}
     \braket{\phi}{\Omega}\sim\int_{\hat{\phi}(\tau=\tau_I)=0}^{\hat{\phi}(\tau=0)=\phi}\mathcal{D}\hat{\phi}\,e^{-S_E[\hat{g},\hat{\phi}]}\,.
     \label{eq:PI_saddle}
\end{equation}
Here the hats indicate that we integrate over field configurations propagating on the saddle point geometry $\hat{g}$, while $S_E$ is the Euclidean action of these fields, with corresponding Hamiltonian $H$. We compute this path integral using the Rindler decomposition \cite{Harlow:2014yka,Nogueira:2021ngh}, i.e.~we split the manifold, without loss of generality, along the $x$-direction into a region $x<0$ and a region $x>0$, called the left and right region. We note that the analytic continuation of the Lorentzian boost generator $K_x$ generates rotations in the Euclidean $\tau x$ plane. Therefore we may evaluate the path integral \eqref{eq:PI_saddle} by rotating by an angle $\pi$ in this plane, instead of integrating from $\tau_I$ to $0$ \cite{Harlow:2014yka}. The rotation is performed around the $x=0$ point at the $\tau=0$ surface. Therefore the field configuration on the right $\phi_R:=\phi(x>0)$ specifies the initial state, while the fields on the left $\phi_L:=\phi(x>0)$ specify the final state. These states belong the left and right Hilbert spaces, which are related by the anti-unitary CPT operator $\Theta$. This operator is also needed to map the state $\bra{\phi}_R$, viewed as a final state from the point of \eqref{eq:PI_saddle}, to an initial state. This leads to the relation $\bra{\phi}_R=\Theta\ket{\phi}_R$, which implies that the boost operator only acts on states in the left Hilbert space \cite{Nogueira:2021ngh}. Thus, the path integral \eqref{eq:PI_saddle} yields
\begin{equation}
    \braket{\phi}{\Omega}=\braket{\phi_L\phi_R}{\Omega}\sim\bra{\phi_L}e^{-\pi K_L}\Theta\ket{\phi_R}\,.
    \label{eq:Rindler_rotation}
\end{equation}
Now consider a set of eigenvectors of $K_L$ denoted by $\ket{E_n}_L$, with eigenvalues $E_n$. In this basis, \eqref{eq:Rindler_rotation} becomes
\begin{equation}
\begin{split}
    \braket{\phi}{\Omega}&\sim\sum\limits_n\bra{\phi_L}e^{-\pi K_L}\Theta\ket{E_n}_L\bra{E_{n,L}}\Theta\ket{\phi_R}\\
    &=\sum\limits_n e^{-\pi E_n}\braket{\phi_L}{E_n}_L\bra{\phi_R}\Theta^\dagger\ket{E_n}_L\,.
\end{split}
\end{equation}
As already mentioned, $\Theta$ is the CPT operator that interchanges the left and the right system, and we may therefore identify $\Theta^\dagger\ket{E_n}_L\sim\ket{E_n}^*_R$ as a state in the right CFT. The proportionality sign $\sim$ indicates that there is a freedom to choose this identification. The canonical choice is an equality, which corresponds to choosing the energy eigenstates in a time-reversal symmetric manner. However, due to the absence of time reversal symmetry in gravity, we may also include phase factors $e^{-i\alpha_n}$, which lead to different identifications of time at the asymptotic boundaries \cite{Papadodimas:2015jra}. From this derivation we find that the Hartle-Hawking state $\ket{\Omega}$ corresponds to the generalized TFD state \eqref{eq:generalised_TFD} at inverse temperature $\beta=2\pi$. 

Now let us consider the overlaps and higher moments of the generalized TFD states from the path integral point of view. The boundary condition for a ket $\tfd_{\gamma}$ corresponds to the lower half of a circle in Euclidean time, with a time shift corresponding to $\gamma$. The boundary condition for a bra $\bra{\text{TFD}}_\alpha$ corresponds to the upper half of a circle in Euclidean time, with a time shift corresponding to $\alpha$. The overlap $\braket{\text{TFD}_\alpha}{\text{TFD}_\gamma}$ is obtained by gluing both halves of the boundary manifold together and filling up the bulk geometry with a Euclidean black hole, as shown in Fig.~\ref{Fig:PI_Black_hole}. This can only be done in a consistent way if the time shifts associated with the bra and the ket are the same. Thus the overlap is given by
\begin{equation}
    \langle\text{TFD}_\alpha\vert\text{TFD}_\gamma\rangle\sim Z_1\,\delta_{\alpha\gamma}\,,
    \label{eq:firs_moment}
\end{equation}
where $Z_1$ is the Gibbons-Hawking partition function of the Euclidean black hole. We choose the normalisation factor for the gravitational path integral \eqref{eq:Hartle-Hawking_state} in such a way that this overlap is equal to a Kronecker delta \eqref{eq:orthogonal}. 

We now consider the second moment \eqref{eq:derivation_overlaps}
\begin{align}
    \label{eq:derivation_second moment}
    &\frac{1}{\mathcal N}\sum\limits_\gamma\vert\langle\text{TFD}_{\alpha}\vert\text{TFD}_\gamma\rangle\vert^2\nonumber\\
    &=\sum\limits_\gamma\sum\limits_{klmn}\delta_{km}\delta_{nl}\,p_kp_lp_mp_ne^{i(\alpha_l-\alpha_k)}e^{i(\gamma_m-\gamma_n)}\nonumber\\
        &=\frac{1}{\mathcal N}\sum\limits_\gamma\sum\limits_{mn}p^2_mp^2_ne^{i(\alpha_n-\alpha_m)}e^{i(\gamma_m-\gamma_n)}\nonumber\\
        &=\frac{\delta_{\alpha\alpha}}{\mathcal N} + \sum\limits_{m}p_m^4= \frac{\delta_{\alpha\alpha}}{\mathcal N}+ \frac{Z(2\beta)}{Z^2(\beta)}=\frac{\delta_{\alpha\alpha}}{\mathcal{N}}+\frac{Z_2}{Z_1}\,,
    \end{align}
    \begin{figure}
    \centering
    \includegraphics[width=1.0\linewidth]{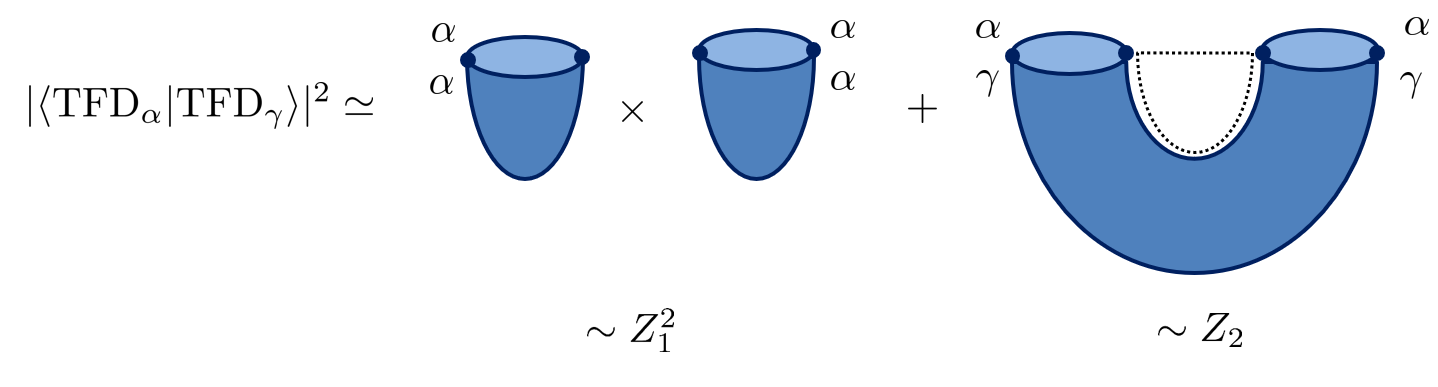} 
        \caption{Leading geometries contributing to the second moment of the Gram matrix \eqref{eq:Gram_matrix}. The first contribution corresponds to two copies of a Euclidean bkack hole, each being proportional to the Gibbons-Hawking partition function $Z_1$. The second contribution corresponds to a wormhole, with partition function $Z_2$. The averaging over the phase shifted states is depicted through the dashed line in this diagram.} 
        \label{Fig:PI_Overlaps}
\end{figure} 
where in the first step we used the definition of the generalised TFD states \eqref{eq:generalised_TFD}, and $p_n=e^{-\beta E_n/2}/\sqrt{Z}$ is the Boltzmann probability, with $Z(\beta)$  the thermal partition function. In the third step in \eqref{eq:derivation_second moment}, we have used the randomness of the phases, in form of the identity $\sum_\gamma e^{i(\gamma_m-\gamma_n)}/\mathcal{N}=\delta_{mn}$, with $\mathcal{N}$ being the total number of states appearing in the sum. The delta arises from the summand, where $\gamma$ is equal to $\alpha$. In the last step we use the fact that the ratio of thermal partition functions is given by the partition function of a replica wormhole with two-throats $Z_2$ divided by $Z_1^2$ \cite{Verlinde:2021kgt}. When calculating the overlap from the gravitational path integral the boundary manifold is given by two copies of a thermal circle times $S^{d-1}$. As previously stated, the path integral in the semiclassical limit corresponds to a sum over classical solutions that are consistent with these boundary conditions. If we only consider planar geometries, which dominate in the $G_N\to0$ and $\Omega\gg 1$ limit, these are given by two copies of the Euclidean black hole, where $\alpha$ and $\gamma$ have to be equal, as well as an Euclidean wormhole connecting both circles (see Fig.~\ref{Fig:PI_Overlaps}).

In general, for the $n$-th moment the boundary manifold is given by $n$ copies of a thermal circle, and we have to consider all planar geometries that are consistent with this boundary condition. The maximally connected geometry is a replica wormhole with $n$ throats, and thus the connected contribution to the $n$-th moment is given by
\begin{equation}
\label{eq:general_overlaps}
\overline{\langle\text{TFD}_{\alpha_1}\vert\text{TFD}_{\alpha_2}\rangle...\langle\text{TFD}_{\alpha_n}\vert\text{TFD}_{\alpha_1}\rangle}_c=\frac{Z(n\beta)}{Z^n(\beta)}=\frac{Z_n}{Z_1^n}\,,
\end{equation}
where $Z_n$ is the partition function of this wormhole. The overline indicates that the n-th moment is obtained from a state averaging procedure. The ratio of partition functions is of order $e^{-S_{\text{BH}}(n-1)}$, again signalling that the wormhole contributions correspond to non-perturbative corrections to the gravitational path integral.

\section{Schwinger-Dyson equation}
\label{apx:Schwinger_Dyson}

Here we derive the eigenvalue density \eqref{eq:result_EV density} of the Gram matrix \eqref{eq:Gram_matrix} using the resolvent method of random matrix theory \cite{CVITANOVIC198149,2009arXiv0911.0087S,Penington:2019kki}. The resolvent is given by 
\begin{equation}
    R_{ij}(\lambda):=\left(\frac{1}{\lambda\mathds{1}-G}\right)_{ij}=\frac{\delta_{ij}}{\lambda}+\sum\limits_{n=1}^{\infty}\frac{(G^n)_{ij}}{\lambda^{n+1}}\,,
    \label{eq:resolvent_matrix}
\end{equation}
and may be represented in a pictorial way as follows. The first term in the sum, which is proportional to the identity matrix may be interpreted as a free propagator, which we represent pictorially by a dashed line, where the open ends correspond to the indices. Contracted indices are represented by closed dashed lines. The Gram matrix is represented by a blue circle, corresponding to a thermal circle, from the point of the gravitational path integral. Thus, \eqref{eq:resolvent_matrix} may be schematically expressed as in the first line of Fig.~\ref{Fig:Schwinger_Dyson}. As explained in the previous section of the supplemental material,  the overlaps between the different microstates, i.e.~the entries of the Gram matrix, correspond to a sum over all possible geometries consistent with the boundary condition given by the number of thermal circles. We work in the semiclassical $G_N\to0$ limit, and consider a large number of microstates $\Omega$. In this limit, only the planar geometries contribute \cite{Penington:2019kki}, which for the first moment just corresponds to a single Euclidean black hole. For the second moment, which is contained in $G_{ij}^2$, these geometries correspond to two copies of a Euclidean black hole, as well as a Euclidean wormhole connecting the two thermal circles, as indicated in the second line of Fig.~\ref{Fig:Schwinger_Dyson}. Therefore, the resolvent corresponds to an infinite sum over all planar geometries, with an arbitrary number of boundaries.

\begin{figure}
    \centering
    \includegraphics[width=1.0\linewidth]{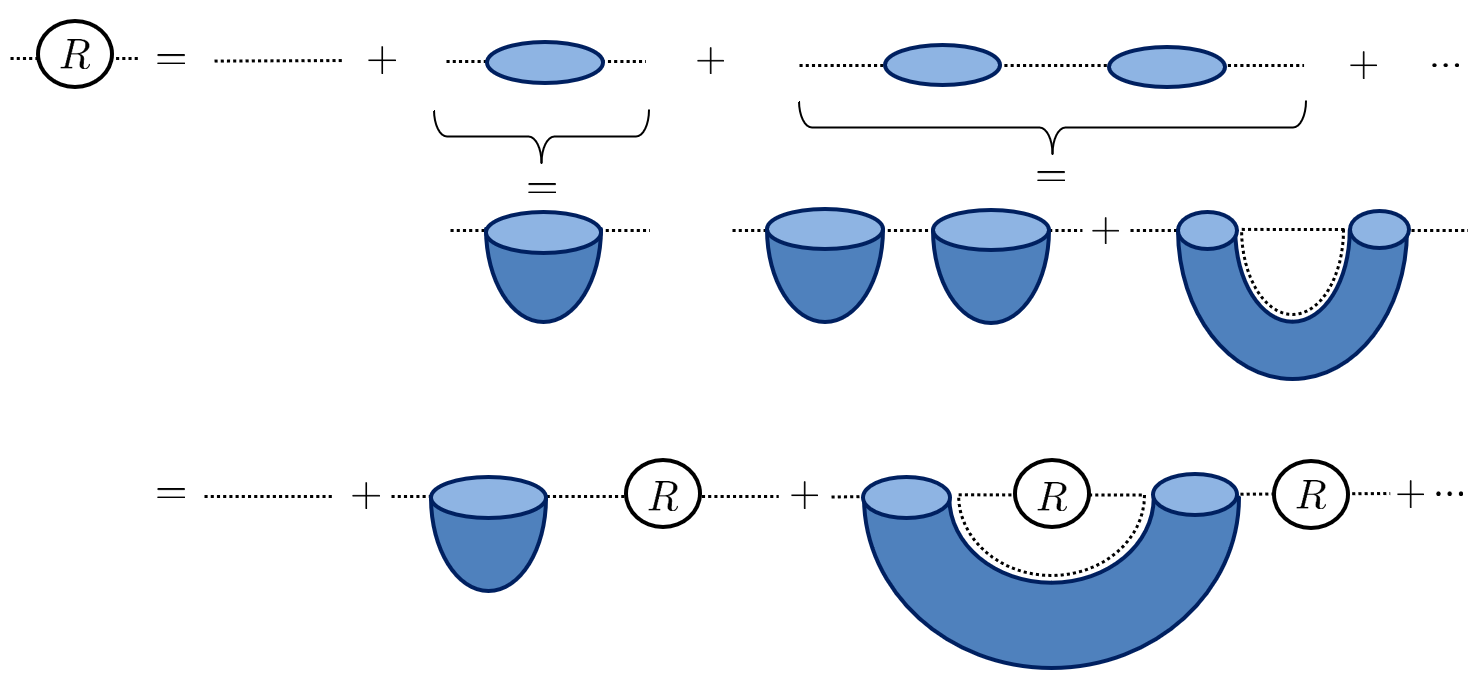} 
        \caption{Diagrammatic representation of the expansion of the resolvent matrix \eqref{eq:resolvent_matrix}. The open dashed lines correspond to free indices $(i,j)$, while internal dashed lines represent contracted indices. The blue circles correspond to an insertion of the Gram matrix \eqref{eq:Gram_matrix}. The second line shows the leading geometries contributing to the overlaps of the Gram matrix. The last line is a reordering of the previous line, as explained in the text, which yields the Schwinger-Dyson equation \eqref{eq:Schwinger_Dyson}.} 
        \label{Fig:Schwinger_Dyson}
\end{figure} 
We reorder this series to obtain a Schwinger-Dyson equation, as indicted in the last line of Fig.~\ref{Fig:Schwinger_Dyson}. The first term is still given by a free propagator, the second term sums all planar geometries in which the first thermal circle is disconnected from all other circles. The third term sums all planar geometries in which the first circle is connected to one other circle via a Euclidean wormhole, and so on. By introducing the trace of the resolvent 
\begin{equation}
    R(\lambda)=\sum\limits_iR_{ii}(\lambda)\,,
    \label{eq:trace_resolvent}
\end{equation}
we find that the pictorial representation is equivalent to the Schwinger-Dyson equation \cite{Penington:2019kki} (see also \cite{Balasubramanian:2022gmo,Balasubramanian:2022lnw,Climent:2024trz,Geng:2024jmm})
 \begin{equation}
    R_{ij}(\lambda)=\frac{\delta_{ij}}{\lambda}+\frac{1}{\lambda}\sum\limits_{n=1}^\infty\frac{Z_n}{Z_1^n}R^{n-1}(\lambda)R_{ij}(\lambda)\,,
    \label{eq:Schwinger_Dyson}
\end{equation}
where $Z_n/Z_1^n$ is the properly normalized partition function of a wormhole with $n$ throats \eqref{eq:general_overlaps}. This partition function arises from the blue geometry appearing in the $n+1$-th term, shown in the last line of Fig.~\ref{Fig:Schwinger_Dyson}. The contracted indices, appearing in the trace \eqref{eq:trace_resolvent} correspond to the closed dashed lines in the pictorial representation in Fig.~\ref{Fig:Schwinger_Dyson}. Taking the trace of \eqref{eq:Schwinger_Dyson} leads to
\begin{equation}
    R(\lambda)=\frac{\Omega}{\lambda}+\frac{1}{\lambda}\sum\limits_{n=1}^\infty\frac{Z_n}{Z_1^n}R^{n}(\lambda)\,.
    \label{eq:trace_Schwinger}
\end{equation}
Note that for the derivation of the overlaps \eqref{eq:general_overlaps} we worked in the canonical ensemble where $Z_n/Z_1^n=Z(n\beta)/Z(\beta)^n$. If we consider a black hole in the canonical ensemble at fixed temperature $T=\beta^{-1}$ the Gibbons-Hawking partition function \cite{PhysRevD.15.2752} is given by
\begin{equation}
    Z(\beta)=e^{-\beta M+ S_{\text{BH}}}=\int\limits_0^\infty dE\,e^{-\beta\,E}z(E)\,,
    \label{eq:laplace_transform}
\end{equation}
where $M$ is the mass of the black hole, and the inverse Laplace transform $z(E)$ may be interpreted as a density of states, which is equal to the microcanonical partition function. As discussed in the main text, in order to properly count the linearly independent microstates, we have to project the states into a microcanoncial energy window $[E_0,E_0+\Delta E]$ \eqref{eq:microcanonical_TFD}. In this ensemble the analog of the wormhole partition function $Z_n$ is given by \cite{Balasubramanian:2022gmo,Penington:2019kki}
\begin{equation}
    \textbf{Z}_n=e^{-n\beta\,E}z(E)\Delta E\,,\quad e^{S_M}=z(E)\Delta E\,,
    \label{eq:microcanical_wormhole}
\end{equation}
where we also introduce the microcanonical entropy 
\begin{equation}
    S_M:=\log(z(E))=S_{\text{BH}}\,.
    \label{eq:microcanical_entropy}
\end{equation}
In the second equality we used the fact that the microcanonical entropy is equal to the Bekenstein-Hawking entropy \eqref{eq:BH_entropy}, which has originally been derived from the canonical partition function \eqref{eq:laplace_transform} \cite{PhysRevD.15.2752}. Consequently, this equality holds due to the equivalence of the ensembles in the $G_N\to0$ limit \cite{Climent:2024trz}. Inserting the microcanonical wormhole partition function \eqref{eq:microcanical_wormhole} into the Schwinger-Dyson equation \eqref{eq:trace_Schwinger} yields
\begin{equation}
    R(\lambda)=\frac{\Omega}{\lambda}+\frac{e^{S_M}}{\lambda}\sum\limits_{n=1}^\infty\left(\frac{R(\lambda)}{e^{S_M}}\right)^n=\frac{\Omega}{\lambda}+\frac{1}{\lambda}\frac{e^{S_M}R(\lambda)}{e^{S_M}-R(\lambda)}\,,
    \label{eq:Solving_schwinger_Dyson}
\end{equation}

\noindent where in the first equality we used $\textbf{Z}_n/\textbf{Z}_1^n=e^{-{S_M}(n-1)}$, which follows directly from the definition of the microcanonical quantities \eqref{eq:microcanical_wormhole}. In the second equality \eqref{eq:Solving_schwinger_Dyson}, we evaluate the geometric series and obtain a quadratic equation for the trace of the resolvent $R(\lambda)$,
\begin{equation}
    \lambda R(\lambda)^2-\left(\lambda e^{S_{\text{BH}}}-\Omega-e^{S_{\text{BH}}}\right) R(\lambda)-\Omega e^{S_{\text{BH}}}=0\,,
    \label{eq:quadratic_equation}
\end{equation}
where we also used the fact that the microcanonical entropy is equal to the Bekenstein-Hawking entropy \eqref{eq:microcanical_entropy}. We solve this quadratic equation for $R(\lambda)$ and obtain the eigenvalue density \eqref{eq:eigenvalue_density} of the Gram matrix \eqref{eq:Gram_matrix} from the discontinuity along the real line as \cite{Eynard2015} 
\begin{equation}
    D(\lambda)=\frac{1}{2\pi i}\lim\limits_{\epsilon\to0}(R(\lambda-i\epsilon)-R(\lambda+i\epsilon))\,,
    \label{eq:branch_cut}
\end{equation}
which yields \eqref{eq:result_EV density}.

\bibliography{literature}

\begin{thebibliography}{65}%
\makeatletter
\providecommand \@ifxundefined [1]{%
 \@ifx{#1\undefined}
}%
\providecommand \@ifnum [1]{%
 \ifnum #1\expandafter \@firstoftwo
 \else \expandafter \@secondoftwo
 \fi
}%
\providecommand \@ifx [1]{%
 \ifx #1\expandafter \@firstoftwo
 \else \expandafter \@secondoftwo
 \fi
}%
\providecommand \natexlab [1]{#1}%
\providecommand \enquote  [1]{``#1''}%
\providecommand \bibnamefont  [1]{#1}%
\providecommand \bibfnamefont [1]{#1}%
\providecommand \citenamefont [1]{#1}%
\providecommand \href@noop [0]{\@secondoftwo}%
\providecommand \href [0]{\begingroup \@sanitize@url \@href}%
\providecommand \@href[1]{\@@startlink{#1}\@@href}%
\providecommand \@@href[1]{\endgroup#1\@@endlink}%
\providecommand \@sanitize@url [0]{\catcode `\\12\catcode `\$12\catcode
  `\&12\catcode `\#12\catcode `\^12\catcode `\_12\catcode `\%12\relax}%
\providecommand \@@startlink[1]{}%
\providecommand \@@endlink[0]{}%
\providecommand \url  [0]{\begingroup\@sanitize@url \@url }%
\providecommand \@url [1]{\endgroup\@href {#1}{\urlprefix }}%
\providecommand \urlprefix  [0]{URL }%
\providecommand \Eprint [0]{\href }%
\providecommand \doibase [0]{http://dx.doi.org/}%
\providecommand \selectlanguage [0]{\@gobble}%
\providecommand \bibinfo  [0]{\@secondoftwo}%
\providecommand \bibfield  [0]{\@secondoftwo}%
\providecommand \translation [1]{[#1]}%
\providecommand \BibitemOpen [0]{}%
\providecommand \bibitemStop [0]{}%
\providecommand \bibitemNoStop [0]{.\EOS\space}%
\providecommand \EOS [0]{\spacefactor3000\relax}%
\providecommand \BibitemShut  [1]{\csname bibitem#1\endcsname}%
\let\auto@bib@innerbib\@empty
\bibitem [{\citenamefont {'t~Hooft}(1993)}]{tHooft:1993dmi}%
  \BibitemOpen
  \bibfield  {author} {\bibinfo {author} {\bibfnamefont {Gerard}\ \bibnamefont
  {'t~Hooft}},\ }\bibfield  {title} {\enquote {\bibinfo {title} {{Dimensional
  reduction in quantum gravity}},}\ }\href@noop {} {\bibfield  {journal}
  {\bibinfo  {journal} {Conf. Proc. C}\ }\textbf {\bibinfo {volume} {930308}},\
  \bibinfo {pages} {284--296} (\bibinfo {year} {1993})},\ \Eprint
  {http://arxiv.org/abs/gr-qc/9310026} {arXiv:gr-qc/9310026} \BibitemShut
  {NoStop}%
\bibitem [{\citenamefont {Susskind}(1995)}]{Susskind:1994vu}%
  \BibitemOpen
  \bibfield  {author} {\bibinfo {author} {\bibfnamefont {Leonard}\ \bibnamefont
  {Susskind}},\ }\bibfield  {title} {\enquote {\bibinfo {title} {{The World as
  a hologram}},}\ }\href {\doibase 10.1063/1.531249} {\bibfield  {journal}
  {\bibinfo  {journal} {J. Math. Phys.}\ }\textbf {\bibinfo {volume} {36}},\
  \bibinfo {pages} {6377--6396} (\bibinfo {year} {1995})},\ \Eprint
  {http://arxiv.org/abs/hep-th/9409089} {arXiv:hep-th/9409089} \BibitemShut
  {NoStop}%
\bibitem [{\citenamefont {Heemskerk}\ \emph {et~al.}(2009)\citenamefont
  {Heemskerk}, \citenamefont {Penedones}, \citenamefont {Polchinski},\ and\
  \citenamefont {Sully}}]{Heemskerk:2009pn}%
  \BibitemOpen
  \bibfield  {author} {\bibinfo {author} {\bibfnamefont {Idse}\ \bibnamefont
  {Heemskerk}}, \bibinfo {author} {\bibfnamefont {Joao}\ \bibnamefont
  {Penedones}}, \bibinfo {author} {\bibfnamefont {Joseph}\ \bibnamefont
  {Polchinski}}, \ and\ \bibinfo {author} {\bibfnamefont {James}\ \bibnamefont
  {Sully}},\ }\bibfield  {title} {\enquote {\bibinfo {title} {{Holography from
  Conformal Field Theory}},}\ }\href {\doibase 10.1088/1126-6708/2009/10/079}
  {\bibfield  {journal} {\bibinfo  {journal} {JHEP}\ }\textbf {\bibinfo
  {volume} {10}},\ \bibinfo {pages} {079} (\bibinfo {year} {2009})},\ \Eprint
  {http://arxiv.org/abs/0907.0151} {arXiv:0907.0151 [hep-th]} \BibitemShut
  {NoStop}%
\bibitem [{\citenamefont {El-Showk}\ and\ \citenamefont
  {Papadodimas}(2012)}]{El-Showk:2011yvt}%
  \BibitemOpen
  \bibfield  {author} {\bibinfo {author} {\bibfnamefont {Sheer}\ \bibnamefont
  {El-Showk}}\ and\ \bibinfo {author} {\bibfnamefont {Kyriakos}\ \bibnamefont
  {Papadodimas}},\ }\bibfield  {title} {\enquote {\bibinfo {title} {{Emergent
  Spacetime and Holographic CFTs}},}\ }\href {\doibase 10.1007/JHEP10(2012)106}
  {\bibfield  {journal} {\bibinfo  {journal} {JHEP}\ }\textbf {\bibinfo
  {volume} {10}},\ \bibinfo {pages} {106} (\bibinfo {year} {2012})},\ \Eprint
  {http://arxiv.org/abs/1101.4163} {arXiv:1101.4163 [hep-th]} \BibitemShut
  {NoStop}%
\bibitem [{Note1()}]{Note1}%
  \BibitemOpen
  \bibinfo {note} {Holographic CFTs are characterized by large central charges
  and sparse spectrum of operators with small conformal dimensions.
  Furthermore, correlators of these low-lying operators in any holographic CFT
  factorize into products of two-point functions, in other words connected
  higher point correlators of these operators vanish \cite
  {El-Showk:2011yvt}.}\BibitemShut {Stop}%
\bibitem [{\citenamefont {Maldacena}(1998)}]{Maldacena:1997re}%
  \BibitemOpen
  \bibfield  {author} {\bibinfo {author} {\bibfnamefont {Juan~Martin}\
  \bibnamefont {Maldacena}},\ }\bibfield  {title} {\enquote {\bibinfo {title}
  {{The Large N limit of superconformal field theories and supergravity}},}\
  }\href {\doibase 10.1023/A:1026654312961} {\bibfield  {journal} {\bibinfo
  {journal} {Adv. Theor. Math. Phys.}\ }\textbf {\bibinfo {volume} {2}},\
  \bibinfo {pages} {231--252} (\bibinfo {year} {1998})},\ \Eprint
  {http://arxiv.org/abs/hep-th/9711200} {arXiv:hep-th/9711200} \BibitemShut
  {NoStop}%
\bibitem [{\citenamefont {Witten}(1998{\natexlab{a}})}]{Witten:1998qj}%
  \BibitemOpen
  \bibfield  {author} {\bibinfo {author} {\bibfnamefont {Edward}\ \bibnamefont
  {Witten}},\ }\bibfield  {title} {\enquote {\bibinfo {title} {{Anti-de Sitter
  space and holography}},}\ }\href {\doibase 10.4310/ATMP.1998.v2.n2.a2}
  {\bibfield  {journal} {\bibinfo  {journal} {Adv. Theor. Math. Phys.}\
  }\textbf {\bibinfo {volume} {2}},\ \bibinfo {pages} {253--291} (\bibinfo
  {year} {1998}{\natexlab{a}})},\ \Eprint {http://arxiv.org/abs/hep-th/9802150}
  {arXiv:hep-th/9802150} \BibitemShut {NoStop}%
\bibitem [{\citenamefont {Gubser}\ \emph {et~al.}(1998)\citenamefont {Gubser},
  \citenamefont {Klebanov},\ and\ \citenamefont {Polyakov}}]{Gubser:1998bc}%
  \BibitemOpen
  \bibfield  {author} {\bibinfo {author} {\bibfnamefont {S.~S.}\ \bibnamefont
  {Gubser}}, \bibinfo {author} {\bibfnamefont {Igor~R.}\ \bibnamefont
  {Klebanov}}, \ and\ \bibinfo {author} {\bibfnamefont {Alexander~M.}\
  \bibnamefont {Polyakov}},\ }\bibfield  {title} {\enquote {\bibinfo {title}
  {{Gauge theory correlators from noncritical string theory}},}\ }\href
  {\doibase 10.1016/S0370-2693(98)00377-3} {\bibfield  {journal} {\bibinfo
  {journal} {Phys. Lett. B}\ }\textbf {\bibinfo {volume} {428}},\ \bibinfo
  {pages} {105--114} (\bibinfo {year} {1998})},\ \Eprint
  {http://arxiv.org/abs/hep-th/9802109} {arXiv:hep-th/9802109} \BibitemShut
  {NoStop}%
\bibitem [{\citenamefont {Witten}(1998{\natexlab{b}})}]{Witten:1998zw}%
  \BibitemOpen
  \bibfield  {author} {\bibinfo {author} {\bibfnamefont {Edward}\ \bibnamefont
  {Witten}},\ }\bibfield  {title} {\enquote {\bibinfo {title} {{Anti-de Sitter
  space, thermal phase transition, and confinement in gauge theories}},}\
  }\href {\doibase 10.4310/ATMP.1998.v2.n3.a3} {\bibfield  {journal} {\bibinfo
  {journal} {Adv. Theor. Math. Phys.}\ }\textbf {\bibinfo {volume} {2}},\
  \bibinfo {pages} {505--532} (\bibinfo {year} {1998}{\natexlab{b}})},\ \Eprint
  {http://arxiv.org/abs/hep-th/9803131} {arXiv:hep-th/9803131} \BibitemShut
  {NoStop}%
\bibitem [{\citenamefont {Chamblin}\ \emph
  {et~al.}(1999{\natexlab{a}})\citenamefont {Chamblin}, \citenamefont
  {Emparan}, \citenamefont {Johnson},\ and\ \citenamefont
  {Myers}}]{Chamblin:1999hg}%
  \BibitemOpen
  \bibfield  {author} {\bibinfo {author} {\bibfnamefont {Andrew}\ \bibnamefont
  {Chamblin}}, \bibinfo {author} {\bibfnamefont {Roberto}\ \bibnamefont
  {Emparan}}, \bibinfo {author} {\bibfnamefont {Clifford~V.}\ \bibnamefont
  {Johnson}}, \ and\ \bibinfo {author} {\bibfnamefont {Robert~C.}\ \bibnamefont
  {Myers}},\ }\bibfield  {title} {\enquote {\bibinfo {title} {{Holography,
  thermodynamics and fluctuations of charged AdS black holes}},}\ }\href
  {\doibase 10.1103/PhysRevD.60.104026} {\bibfield  {journal} {\bibinfo
  {journal} {Phys. Rev. D}\ }\textbf {\bibinfo {volume} {60}},\ \bibinfo
  {pages} {104026} (\bibinfo {year} {1999}{\natexlab{a}})},\ \Eprint
  {http://arxiv.org/abs/hep-th/9904197} {arXiv:hep-th/9904197} \BibitemShut
  {NoStop}%
\bibitem [{\citenamefont {Chamblin}\ \emph
  {et~al.}(1999{\natexlab{b}})\citenamefont {Chamblin}, \citenamefont
  {Emparan}, \citenamefont {Johnson},\ and\ \citenamefont
  {Myers}}]{Chamblin:1999tk}%
  \BibitemOpen
  \bibfield  {author} {\bibinfo {author} {\bibfnamefont {Andrew}\ \bibnamefont
  {Chamblin}}, \bibinfo {author} {\bibfnamefont {Roberto}\ \bibnamefont
  {Emparan}}, \bibinfo {author} {\bibfnamefont {Clifford~V.}\ \bibnamefont
  {Johnson}}, \ and\ \bibinfo {author} {\bibfnamefont {Robert~C.}\ \bibnamefont
  {Myers}},\ }\bibfield  {title} {\enquote {\bibinfo {title} {{Charged AdS
  black holes and catastrophic holography}},}\ }\href {\doibase
  10.1103/PhysRevD.60.064018} {\bibfield  {journal} {\bibinfo  {journal} {Phys.
  Rev. D}\ }\textbf {\bibinfo {volume} {60}},\ \bibinfo {pages} {064018}
  (\bibinfo {year} {1999}{\natexlab{b}})},\ \Eprint
  {http://arxiv.org/abs/hep-th/9902170} {arXiv:hep-th/9902170} \BibitemShut
  {NoStop}%
\bibitem [{\citenamefont {Ryu}\ and\ \citenamefont
  {Takayanagi}(2006{\natexlab{a}})}]{Ryu:2006bv}%
  \BibitemOpen
  \bibfield  {author} {\bibinfo {author} {\bibfnamefont {Shinsei}\ \bibnamefont
  {Ryu}}\ and\ \bibinfo {author} {\bibfnamefont {Tadashi}\ \bibnamefont
  {Takayanagi}},\ }\bibfield  {title} {\enquote {\bibinfo {title} {{Holographic
  derivation of entanglement entropy from AdS/CFT}},}\ }\href {\doibase
  10.1103/PhysRevLett.96.181602} {\bibfield  {journal} {\bibinfo  {journal}
  {Phys. Rev. Lett.}\ }\textbf {\bibinfo {volume} {96}},\ \bibinfo {pages}
  {181602} (\bibinfo {year} {2006}{\natexlab{a}})},\ \Eprint
  {http://arxiv.org/abs/hep-th/0603001} {arXiv:hep-th/0603001} \BibitemShut
  {NoStop}%
\bibitem [{\citenamefont {Ryu}\ and\ \citenamefont
  {Takayanagi}(2006{\natexlab{b}})}]{Ryu:2006ef}%
  \BibitemOpen
  \bibfield  {author} {\bibinfo {author} {\bibfnamefont {Shinsei}\ \bibnamefont
  {Ryu}}\ and\ \bibinfo {author} {\bibfnamefont {Tadashi}\ \bibnamefont
  {Takayanagi}},\ }\bibfield  {title} {\enquote {\bibinfo {title} {{Aspects of
  Holographic Entanglement Entropy}},}\ }\href {\doibase
  10.1088/1126-6708/2006/08/045} {\bibfield  {journal} {\bibinfo  {journal}
  {JHEP}\ }\textbf {\bibinfo {volume} {08}},\ \bibinfo {pages} {045} (\bibinfo
  {year} {2006}{\natexlab{b}})},\ \Eprint {http://arxiv.org/abs/hep-th/0605073}
  {arXiv:hep-th/0605073} \BibitemShut {NoStop}%
\bibitem [{\citenamefont {Hubeny}\ \emph {et~al.}(2007)\citenamefont {Hubeny},
  \citenamefont {Rangamani},\ and\ \citenamefont {Takayanagi}}]{Hubeny:2007xt}%
  \BibitemOpen
  \bibfield  {author} {\bibinfo {author} {\bibfnamefont {Veronika~E.}\
  \bibnamefont {Hubeny}}, \bibinfo {author} {\bibfnamefont {Mukund}\
  \bibnamefont {Rangamani}}, \ and\ \bibinfo {author} {\bibfnamefont {Tadashi}\
  \bibnamefont {Takayanagi}},\ }\bibfield  {title} {\enquote {\bibinfo {title}
  {{A Covariant holographic entanglement entropy proposal}},}\ }\href {\doibase
  10.1088/1126-6708/2007/07/062} {\bibfield  {journal} {\bibinfo  {journal}
  {JHEP}\ }\textbf {\bibinfo {volume} {07}},\ \bibinfo {pages} {062} (\bibinfo
  {year} {2007})},\ \Eprint {http://arxiv.org/abs/0705.0016} {arXiv:0705.0016
  [hep-th]} \BibitemShut {NoStop}%
\bibitem [{\citenamefont {Akers}\ \emph {et~al.}(2024)\citenamefont {Akers},
  \citenamefont {Engelhardt}, \citenamefont {Harlow}, \citenamefont
  {Penington},\ and\ \citenamefont {Vardhan}}]{Akers:2022qdl}%
  \BibitemOpen
  \bibfield  {author} {\bibinfo {author} {\bibfnamefont {Chris}\ \bibnamefont
  {Akers}}, \bibinfo {author} {\bibfnamefont {Netta}\ \bibnamefont
  {Engelhardt}}, \bibinfo {author} {\bibfnamefont {Daniel}\ \bibnamefont
  {Harlow}}, \bibinfo {author} {\bibfnamefont {Geoff}\ \bibnamefont
  {Penington}}, \ and\ \bibinfo {author} {\bibfnamefont {Shreya}\ \bibnamefont
  {Vardhan}},\ }\bibfield  {title} {\enquote {\bibinfo {title} {{The black hole
  interior from non-isometric codes and complexity}},}\ }\href {\doibase
  10.1007/JHEP06(2024)155} {\bibfield  {journal} {\bibinfo  {journal} {JHEP}\
  }\textbf {\bibinfo {volume} {06}},\ \bibinfo {pages} {155} (\bibinfo {year}
  {2024})},\ \Eprint {http://arxiv.org/abs/2207.06536} {arXiv:2207.06536
  [hep-th]} \BibitemShut {NoStop}%
\bibitem [{\citenamefont {Faulkner}\ and\ \citenamefont
  {Li}(2022)}]{Faulkner:2022ada}%
  \BibitemOpen
  \bibfield  {author} {\bibinfo {author} {\bibfnamefont {Thomas}\ \bibnamefont
  {Faulkner}}\ and\ \bibinfo {author} {\bibfnamefont {Min}\ \bibnamefont
  {Li}},\ }\bibfield  {title} {\enquote {\bibinfo {title} {{Asymptotically
  isometric codes for holography}},}\ }\href@noop {} {\  (\bibinfo {year}
  {2022})},\ \Eprint {http://arxiv.org/abs/2211.12439} {arXiv:2211.12439
  [hep-th]} \BibitemShut {NoStop}%
\bibitem [{\citenamefont {DeWolfe}\ and\ \citenamefont
  {Higginbotham}(2023)}]{DeWolfe:2023iuq}%
  \BibitemOpen
  \bibfield  {author} {\bibinfo {author} {\bibfnamefont {Oliver}\ \bibnamefont
  {DeWolfe}}\ and\ \bibinfo {author} {\bibfnamefont {Kenneth}\ \bibnamefont
  {Higginbotham}},\ }\bibfield  {title} {\enquote {\bibinfo {title}
  {{Non-isometric codes for the black hole interior from fundamental and
  effective dynamics}},}\ }\href {\doibase 10.1007/JHEP09(2023)068} {\bibfield
  {journal} {\bibinfo  {journal} {JHEP}\ }\textbf {\bibinfo {volume} {09}},\
  \bibinfo {pages} {068} (\bibinfo {year} {2023})},\ \Eprint
  {http://arxiv.org/abs/2304.12345} {arXiv:2304.12345 [hep-th]} \BibitemShut
  {NoStop}%
\bibitem [{\citenamefont {Antonini}\ \emph {et~al.}(2024)\citenamefont
  {Antonini}, \citenamefont {Balasubramanian}, \citenamefont {Bao},
  \citenamefont {Cao},\ and\ \citenamefont {Chemissany}}]{Antonini:2024yif}%
  \BibitemOpen
  \bibfield  {author} {\bibinfo {author} {\bibfnamefont {Stefano}\ \bibnamefont
  {Antonini}}, \bibinfo {author} {\bibfnamefont {Vijay}\ \bibnamefont
  {Balasubramanian}}, \bibinfo {author} {\bibfnamefont {Ning}\ \bibnamefont
  {Bao}}, \bibinfo {author} {\bibfnamefont {ChunJun}\ \bibnamefont {Cao}}, \
  and\ \bibinfo {author} {\bibfnamefont {Wissam}\ \bibnamefont {Chemissany}},\
  }\bibfield  {title} {\enquote {\bibinfo {title} {{Non-isometry,
  State-Dependence and Holography}},}\ }\href@noop {} {\  (\bibinfo {year}
  {2024})},\ \Eprint {http://arxiv.org/abs/2411.07296} {arXiv:2411.07296
  [hep-th]} \BibitemShut {NoStop}%
\bibitem [{\citenamefont {Harlow}\ and\ \citenamefont
  {Jafferis}(2020)}]{Harlow:2018tqv}%
  \BibitemOpen
  \bibfield  {author} {\bibinfo {author} {\bibfnamefont {Daniel}\ \bibnamefont
  {Harlow}}\ and\ \bibinfo {author} {\bibfnamefont {Daniel}\ \bibnamefont
  {Jafferis}},\ }\bibfield  {title} {\enquote {\bibinfo {title} {{The
  Factorization Problem in Jackiw-Teitelboim Gravity}},}\ }\href {\doibase
  10.1007/JHEP02(2020)177} {\bibfield  {journal} {\bibinfo  {journal} {JHEP}\
  }\textbf {\bibinfo {volume} {02}},\ \bibinfo {pages} {177} (\bibinfo {year}
  {2020})},\ \Eprint {http://arxiv.org/abs/1804.01081} {arXiv:1804.01081
  [hep-th]} \BibitemShut {NoStop}%
\bibitem [{\citenamefont {Jafferis}\ and\ \citenamefont
  {Kolchmeyer}(2019)}]{Jafferis:2019wkd}%
  \BibitemOpen
  \bibfield  {author} {\bibinfo {author} {\bibfnamefont {Daniel~Louis}\
  \bibnamefont {Jafferis}}\ and\ \bibinfo {author} {\bibfnamefont {David~K.}\
  \bibnamefont {Kolchmeyer}},\ }\bibfield  {title} {\enquote {\bibinfo {title}
  {{Entanglement Entropy in Jackiw-Teitelboim Gravity}},}\ }\href@noop {} {\
  (\bibinfo {year} {2019})},\ \Eprint {http://arxiv.org/abs/1911.10663}
  {arXiv:1911.10663 [hep-th]} \BibitemShut {NoStop}%
\bibitem [{\citenamefont {Maldacena}\ and\ \citenamefont
  {Susskind}(2013)}]{Maldacena:2013xja}%
  \BibitemOpen
  \bibfield  {author} {\bibinfo {author} {\bibfnamefont {Juan}\ \bibnamefont
  {Maldacena}}\ and\ \bibinfo {author} {\bibfnamefont {Leonard}\ \bibnamefont
  {Susskind}},\ }\bibfield  {title} {\enquote {\bibinfo {title} {{Cool horizons
  for entangled black holes}},}\ }\href {\doibase 10.1002/prop.201300020}
  {\bibfield  {journal} {\bibinfo  {journal} {Fortsch. Phys.}\ }\textbf
  {\bibinfo {volume} {61}},\ \bibinfo {pages} {781--811} (\bibinfo {year}
  {2013})},\ \Eprint {http://arxiv.org/abs/1306.0533} {arXiv:1306.0533
  [hep-th]} \BibitemShut {NoStop}%
\bibitem [{\citenamefont {Van~Raamsdonk}(2010)}]{VanRaamsdonk:2010pw}%
  \BibitemOpen
  \bibfield  {author} {\bibinfo {author} {\bibfnamefont {Mark}\ \bibnamefont
  {Van~Raamsdonk}},\ }\bibfield  {title} {\enquote {\bibinfo {title} {{Building
  up spacetime with quantum entanglement}},}\ }\href {\doibase
  10.1142/S0218271810018529} {\bibfield  {journal} {\bibinfo  {journal} {Gen.
  Rel. Grav.}\ }\textbf {\bibinfo {volume} {42}},\ \bibinfo {pages}
  {2323--2329} (\bibinfo {year} {2010})},\ \Eprint
  {http://arxiv.org/abs/1005.3035} {arXiv:1005.3035 [hep-th]} \BibitemShut
  {NoStop}%
\bibitem [{\citenamefont {Maldacena}(2003)}]{Maldacena:2001kr}%
  \BibitemOpen
  \bibfield  {author} {\bibinfo {author} {\bibfnamefont {Juan~Martin}\
  \bibnamefont {Maldacena}},\ }\bibfield  {title} {\enquote {\bibinfo {title}
  {{Eternal black holes in anti-de Sitter}},}\ }\href {\doibase
  10.1088/1126-6708/2003/04/021} {\bibfield  {journal} {\bibinfo  {journal}
  {JHEP}\ }\textbf {\bibinfo {volume} {04}},\ \bibinfo {pages} {021} (\bibinfo
  {year} {2003})},\ \Eprint {http://arxiv.org/abs/hep-th/0106112}
  {arXiv:hep-th/0106112} \BibitemShut {NoStop}%
\bibitem [{\citenamefont {Haag}(1996)}]{Book:Haag}%
  \BibitemOpen
  \bibfield  {author} {\bibinfo {author} {\bibfnamefont {Rudolf}\ \bibnamefont
  {Haag}},\ }\href@noop {} {\emph {\bibinfo {title} {Local Quantum Physics:
  Fields, Particles, Algebras}}}\ (\bibinfo  {publisher} {Springer},\ \bibinfo
  {year} {1996})\BibitemShut {NoStop}%
\bibitem [{\citenamefont {Witten}(2022)}]{Witten:2021unn}%
  \BibitemOpen
  \bibfield  {author} {\bibinfo {author} {\bibfnamefont {Edward}\ \bibnamefont
  {Witten}},\ }\bibfield  {title} {\enquote {\bibinfo {title} {{Gravity and the
  crossed product}},}\ }\href {\doibase 10.1007/JHEP10(2022)008} {\bibfield
  {journal} {\bibinfo  {journal} {JHEP}\ }\textbf {\bibinfo {volume} {10}},\
  \bibinfo {pages} {008} (\bibinfo {year} {2022})},\ \Eprint
  {http://arxiv.org/abs/2112.12828} {arXiv:2112.12828 [hep-th]} \BibitemShut
  {NoStop}%
\bibitem [{\citenamefont {Papadodimas}\ and\ \citenamefont
  {Raju}(2015)}]{Papadodimas:2015xma}%
  \BibitemOpen
  \bibfield  {author} {\bibinfo {author} {\bibfnamefont {Kyriakos}\
  \bibnamefont {Papadodimas}}\ and\ \bibinfo {author} {\bibfnamefont {Suvrat}\
  \bibnamefont {Raju}},\ }\bibfield  {title} {\enquote {\bibinfo {title}
  {{Local Operators in the Eternal Black Hole}},}\ }\href {\doibase
  10.1103/PhysRevLett.115.211601} {\bibfield  {journal} {\bibinfo  {journal}
  {Phys. Rev. Lett.}\ }\textbf {\bibinfo {volume} {115}},\ \bibinfo {pages}
  {211601} (\bibinfo {year} {2015})},\ \Eprint
  {http://arxiv.org/abs/1502.06692} {arXiv:1502.06692 [hep-th]} \BibitemShut
  {NoStop}%
\bibitem [{\citenamefont {Papadodimas}\ and\ \citenamefont
  {Raju}(2016)}]{Papadodimas:2015jra}%
  \BibitemOpen
  \bibfield  {author} {\bibinfo {author} {\bibfnamefont {Kyriakos}\
  \bibnamefont {Papadodimas}}\ and\ \bibinfo {author} {\bibfnamefont {Suvrat}\
  \bibnamefont {Raju}},\ }\bibfield  {title} {\enquote {\bibinfo {title}
  {{Remarks on the necessity and implications of state-dependence in the black
  hole interior}},}\ }\href {\doibase 10.1103/PhysRevD.93.084049} {\bibfield
  {journal} {\bibinfo  {journal} {Phys. Rev. D}\ }\textbf {\bibinfo {volume}
  {93}},\ \bibinfo {pages} {084049} (\bibinfo {year} {2016})},\ \Eprint
  {http://arxiv.org/abs/1503.08825} {arXiv:1503.08825 [hep-th]} \BibitemShut
  {NoStop}%
\bibitem [{\citenamefont {Verlinde}(2020)}]{Verlinde:2020upt}%
  \BibitemOpen
  \bibfield  {author} {\bibinfo {author} {\bibfnamefont {Herman}\ \bibnamefont
  {Verlinde}},\ }\bibfield  {title} {\enquote {\bibinfo {title} {{ER = EPR
  revisited: On the Entropy of an Einstein-Rosen Bridge}},}\ }\href@noop {} {\
  (\bibinfo {year} {2020})},\ \Eprint {http://arxiv.org/abs/2003.13117}
  {arXiv:2003.13117 [hep-th]} \BibitemShut {NoStop}%
\bibitem [{\citenamefont {Nogueira}\ \emph {et~al.}(2022)\citenamefont
  {Nogueira}, \citenamefont {Banerjee}, \citenamefont {Dorband}, \citenamefont
  {Meyer}, \citenamefont {Brink},\ and\ \citenamefont
  {Erdmenger}}]{Nogueira:2021ngh}%
  \BibitemOpen
  \bibfield  {author} {\bibinfo {author} {\bibfnamefont {Flavio~S.}\
  \bibnamefont {Nogueira}}, \bibinfo {author} {\bibfnamefont {Souvik}\
  \bibnamefont {Banerjee}}, \bibinfo {author} {\bibfnamefont {Moritz}\
  \bibnamefont {Dorband}}, \bibinfo {author} {\bibfnamefont {Ren\'e}\
  \bibnamefont {Meyer}}, \bibinfo {author} {\bibfnamefont {Jeroen van~den}\
  \bibnamefont {Brink}}, \ and\ \bibinfo {author} {\bibfnamefont {Johanna}\
  \bibnamefont {Erdmenger}},\ }\bibfield  {title} {\enquote {\bibinfo {title}
  {{Geometric phases distinguish entangled states in wormhole quantum
  mechanics}},}\ }\href {\doibase 10.1103/PhysRevD.105.L081903} {\bibfield
  {journal} {\bibinfo  {journal} {Phys. Rev. D}\ }\textbf {\bibinfo {volume}
  {105}},\ \bibinfo {pages} {L081903} (\bibinfo {year} {2022})},\ \Eprint
  {http://arxiv.org/abs/2109.06190} {arXiv:2109.06190 [hep-th]} \BibitemShut
  {NoStop}%
\bibitem [{\citenamefont {Banerjee}\ \emph {et~al.}(2022)\citenamefont
  {Banerjee}, \citenamefont {Dorband}, \citenamefont {Erdmenger}, \citenamefont
  {Meyer},\ and\ \citenamefont {Weigel}}]{Banerjee:2022jnv}%
  \BibitemOpen
  \bibfield  {author} {\bibinfo {author} {\bibfnamefont {Souvik}\ \bibnamefont
  {Banerjee}}, \bibinfo {author} {\bibfnamefont {Moritz}\ \bibnamefont
  {Dorband}}, \bibinfo {author} {\bibfnamefont {Johanna}\ \bibnamefont
  {Erdmenger}}, \bibinfo {author} {\bibfnamefont {Ren\'e}\ \bibnamefont
  {Meyer}}, \ and\ \bibinfo {author} {\bibfnamefont {Anna-Lena}\ \bibnamefont
  {Weigel}},\ }\bibfield  {title} {\enquote {\bibinfo {title} {{Berry phases,
  wormholes and factorization in AdS/CFT}},}\ }\href {\doibase
  10.1007/JHEP08(2022)162} {\bibfield  {journal} {\bibinfo  {journal} {JHEP}\
  }\textbf {\bibinfo {volume} {08}},\ \bibinfo {pages} {162} (\bibinfo {year}
  {2022})},\ \Eprint {http://arxiv.org/abs/2202.11717} {arXiv:2202.11717
  [hep-th]} \BibitemShut {NoStop}%
\bibitem [{\citenamefont {Banerjee}\ \emph
  {et~al.}(2023{\natexlab{a}})\citenamefont {Banerjee}, \citenamefont
  {Basteiro}, \citenamefont {Das},\ and\ \citenamefont
  {Dorband}}]{Banerjee:2023liw}%
  \BibitemOpen
  \bibfield  {author} {\bibinfo {author} {\bibfnamefont {Souvik}\ \bibnamefont
  {Banerjee}}, \bibinfo {author} {\bibfnamefont {Pablo}\ \bibnamefont
  {Basteiro}}, \bibinfo {author} {\bibfnamefont {Rathindra~Nath}\ \bibnamefont
  {Das}}, \ and\ \bibinfo {author} {\bibfnamefont {Moritz}\ \bibnamefont
  {Dorband}},\ }\bibfield  {title} {\enquote {\bibinfo {title} {{Geometric
  Quantum Discord Signals Non-Factorization}},}\ }\href@noop {} {\  (\bibinfo
  {year} {2023}{\natexlab{a}})},\ \Eprint {http://arxiv.org/abs/2305.04952}
  {arXiv:2305.04952 [hep-th]} \BibitemShut {NoStop}%
\bibitem [{\citenamefont {Banerjee}\ \emph
  {et~al.}(2023{\natexlab{b}})\citenamefont {Banerjee}, \citenamefont
  {Dorband}, \citenamefont {Erdmenger},\ and\ \citenamefont
  {Weigel}}]{Banerjee:2023eew}%
  \BibitemOpen
  \bibfield  {author} {\bibinfo {author} {\bibfnamefont {Souvik}\ \bibnamefont
  {Banerjee}}, \bibinfo {author} {\bibfnamefont {Moritz}\ \bibnamefont
  {Dorband}}, \bibinfo {author} {\bibfnamefont {Johanna}\ \bibnamefont
  {Erdmenger}}, \ and\ \bibinfo {author} {\bibfnamefont {Anna-Lena}\
  \bibnamefont {Weigel}},\ }\bibfield  {title} {\enquote {\bibinfo {title}
  {{Geometric phases characterise operator algebras and missing
  information}},}\ }\href {\doibase 10.1007/JHEP10(2023)026} {\bibfield
  {journal} {\bibinfo  {journal} {JHEP}\ }\textbf {\bibinfo {volume} {10}},\
  \bibinfo {pages} {026} (\bibinfo {year} {2023}{\natexlab{b}})},\ \Eprint
  {http://arxiv.org/abs/2306.00055} {arXiv:2306.00055 [hep-th]} \BibitemShut
  {NoStop}%
\bibitem [{\citenamefont {Penington}\ \emph {et~al.}(2022)\citenamefont
  {Penington}, \citenamefont {Shenker}, \citenamefont {Stanford},\ and\
  \citenamefont {Yang}}]{Penington:2019kki}%
  \BibitemOpen
  \bibfield  {author} {\bibinfo {author} {\bibfnamefont {Geoff}\ \bibnamefont
  {Penington}}, \bibinfo {author} {\bibfnamefont {Stephen~H.}\ \bibnamefont
  {Shenker}}, \bibinfo {author} {\bibfnamefont {Douglas}\ \bibnamefont
  {Stanford}}, \ and\ \bibinfo {author} {\bibfnamefont {Zhenbin}\ \bibnamefont
  {Yang}},\ }\bibfield  {title} {\enquote {\bibinfo {title} {{Replica wormholes
  and the black hole interior}},}\ }\href {\doibase 10.1007/JHEP03(2022)205}
  {\bibfield  {journal} {\bibinfo  {journal} {JHEP}\ }\textbf {\bibinfo
  {volume} {03}},\ \bibinfo {pages} {205} (\bibinfo {year} {2022})},\ \Eprint
  {http://arxiv.org/abs/1911.11977} {arXiv:1911.11977 [hep-th]} \BibitemShut
  {NoStop}%
\bibitem [{\citenamefont {Brown}\ \emph {et~al.}(2020)\citenamefont {Brown},
  \citenamefont {Gharibyan}, \citenamefont {Penington},\ and\ \citenamefont
  {Susskind}}]{Brown:2019rox}%
  \BibitemOpen
  \bibfield  {author} {\bibinfo {author} {\bibfnamefont {Adam~R.}\ \bibnamefont
  {Brown}}, \bibinfo {author} {\bibfnamefont {Hrant}\ \bibnamefont
  {Gharibyan}}, \bibinfo {author} {\bibfnamefont {Geoff}\ \bibnamefont
  {Penington}}, \ and\ \bibinfo {author} {\bibfnamefont {Leonard}\ \bibnamefont
  {Susskind}},\ }\bibfield  {title} {\enquote {\bibinfo {title} {{The
  Python\textquoteright{}s Lunch: geometric obstructions to decoding Hawking
  radiation}},}\ }\href {\doibase 10.1007/JHEP08(2020)121} {\bibfield
  {journal} {\bibinfo  {journal} {JHEP}\ }\textbf {\bibinfo {volume} {08}},\
  \bibinfo {pages} {121} (\bibinfo {year} {2020})},\ \Eprint
  {http://arxiv.org/abs/1912.00228} {arXiv:1912.00228 [hep-th]} \BibitemShut
  {NoStop}%
\bibitem [{\citenamefont {Geng}\ and\ \citenamefont
  {Jiang}(2024)}]{Geng:2024jmm}%
  \BibitemOpen
  \bibfield  {author} {\bibinfo {author} {\bibfnamefont {Hao}\ \bibnamefont
  {Geng}}\ and\ \bibinfo {author} {\bibfnamefont {Yikun}\ \bibnamefont
  {Jiang}},\ }\bibfield  {title} {\enquote {\bibinfo {title} {{Microscopic
  Origin of the Entropy of Single-sided Black Holes}},}\ }\href@noop {} {\
  (\bibinfo {year} {2024})},\ \Eprint {http://arxiv.org/abs/2409.12219}
  {arXiv:2409.12219 [hep-th]} \BibitemShut {NoStop}%
\bibitem [{\citenamefont {Balasubramanian}\ \emph
  {et~al.}(2022{\natexlab{a}})\citenamefont {Balasubramanian}, \citenamefont
  {Lawrence}, \citenamefont {Magan},\ and\ \citenamefont
  {Sasieta}}]{Balasubramanian:2022gmo}%
  \BibitemOpen
  \bibfield  {author} {\bibinfo {author} {\bibfnamefont {Vijay}\ \bibnamefont
  {Balasubramanian}}, \bibinfo {author} {\bibfnamefont {Albion}\ \bibnamefont
  {Lawrence}}, \bibinfo {author} {\bibfnamefont {Javier~M.}\ \bibnamefont
  {Magan}}, \ and\ \bibinfo {author} {\bibfnamefont {Martin}\ \bibnamefont
  {Sasieta}},\ }\bibfield  {title} {\enquote {\bibinfo {title} {{Microscopic
  origin of the entropy of black holes in general relativity}},}\ }\href@noop
  {} {\  (\bibinfo {year} {2022}{\natexlab{a}})},\ \Eprint
  {http://arxiv.org/abs/2212.02447} {arXiv:2212.02447 [hep-th]} \BibitemShut
  {NoStop}%
\bibitem [{\citenamefont {Balasubramanian}\ \emph
  {et~al.}(2022{\natexlab{b}})\citenamefont {Balasubramanian}, \citenamefont
  {Lawrence}, \citenamefont {Magan},\ and\ \citenamefont
  {Sasieta}}]{Balasubramanian:2022lnw}%
  \BibitemOpen
  \bibfield  {author} {\bibinfo {author} {\bibfnamefont {Vijay}\ \bibnamefont
  {Balasubramanian}}, \bibinfo {author} {\bibfnamefont {Albion}\ \bibnamefont
  {Lawrence}}, \bibinfo {author} {\bibfnamefont {Javier~M.}\ \bibnamefont
  {Magan}}, \ and\ \bibinfo {author} {\bibfnamefont {Martin}\ \bibnamefont
  {Sasieta}},\ }\bibfield  {title} {\enquote {\bibinfo {title} {{Microscopic
  origin of the entropy of astrophysical black holes}},}\ }\href@noop {} {\
  (\bibinfo {year} {2022}{\natexlab{b}})},\ \Eprint
  {http://arxiv.org/abs/2212.08623} {arXiv:2212.08623 [hep-th]} \BibitemShut
  {NoStop}%
\bibitem [{\citenamefont {Climent}\ \emph {et~al.}(2024)\citenamefont
  {Climent}, \citenamefont {Emparan}, \citenamefont {Magan}, \citenamefont
  {Sasieta},\ and\ \citenamefont {Vilar~L\'opez}}]{Climent:2024trz}%
  \BibitemOpen
  \bibfield  {author} {\bibinfo {author} {\bibfnamefont {Ana}\ \bibnamefont
  {Climent}}, \bibinfo {author} {\bibfnamefont {Roberto}\ \bibnamefont
  {Emparan}}, \bibinfo {author} {\bibfnamefont {Javier~M.}\ \bibnamefont
  {Magan}}, \bibinfo {author} {\bibfnamefont {Martin}\ \bibnamefont {Sasieta}},
  \ and\ \bibinfo {author} {\bibfnamefont {Alejandro}\ \bibnamefont
  {Vilar~L\'opez}},\ }\bibfield  {title} {\enquote {\bibinfo {title}
  {{Universal construction of black hole microstates}},}\ }\href {\doibase
  10.1103/PhysRevD.109.086024} {\bibfield  {journal} {\bibinfo  {journal}
  {Phys. Rev. D}\ }\textbf {\bibinfo {volume} {109}},\ \bibinfo {pages}
  {086024} (\bibinfo {year} {2024})},\ \Eprint
  {http://arxiv.org/abs/2401.08775} {arXiv:2401.08775 [hep-th]} \BibitemShut
  {NoStop}%
\bibitem [{\citenamefont {Boruch}\ \emph {et~al.}(2024)\citenamefont {Boruch},
  \citenamefont {Iliesiu}, \citenamefont {Lin},\ and\ \citenamefont
  {Yan}}]{Boruch:2024kvv}%
  \BibitemOpen
  \bibfield  {author} {\bibinfo {author} {\bibfnamefont {Jan}\ \bibnamefont
  {Boruch}}, \bibinfo {author} {\bibfnamefont {Luca~V.}\ \bibnamefont
  {Iliesiu}}, \bibinfo {author} {\bibfnamefont {Guanda}\ \bibnamefont {Lin}}, \
  and\ \bibinfo {author} {\bibfnamefont {Cynthia}\ \bibnamefont {Yan}},\
  }\bibfield  {title} {\enquote {\bibinfo {title} {{How the Hilbert space of
  two-sided black holes factorises}},}\ }\href@noop {} {\  (\bibinfo {year}
  {2024})},\ \Eprint {http://arxiv.org/abs/2406.04396} {arXiv:2406.04396
  [hep-th]} \BibitemShut {NoStop}%
\bibitem [{\citenamefont {Balasubramanian}\ \emph {et~al.}(2024)\citenamefont
  {Balasubramanian}, \citenamefont {Craps}, \citenamefont {Hernandez},
  \citenamefont {Khramtsov},\ and\ \citenamefont
  {Knysh}}]{Balasubramanian:2024yxk}%
  \BibitemOpen
  \bibfield  {author} {\bibinfo {author} {\bibfnamefont {Vijay}\ \bibnamefont
  {Balasubramanian}}, \bibinfo {author} {\bibfnamefont {Ben}\ \bibnamefont
  {Craps}}, \bibinfo {author} {\bibfnamefont {Juan}\ \bibnamefont {Hernandez}},
  \bibinfo {author} {\bibfnamefont {Mikhail}\ \bibnamefont {Khramtsov}}, \ and\
  \bibinfo {author} {\bibfnamefont {Maria}\ \bibnamefont {Knysh}},\ }\bibfield
  {title} {\enquote {\bibinfo {title} {{Factorization of the Hilbert space of
  eternal black holes in general relativity}},}\ }\href@noop {} {\  (\bibinfo
  {year} {2024})},\ \Eprint {http://arxiv.org/abs/2410.00091} {arXiv:2410.00091
  [hep-th]} \BibitemShut {NoStop}%
\bibitem [{\citenamefont {Li}(2024)}]{Li:2024nft}%
  \BibitemOpen
  \bibfield  {author} {\bibinfo {author} {\bibfnamefont {Pan}\ \bibnamefont
  {Li}},\ }\bibfield  {title} {\enquote {\bibinfo {title} {{Notes on the
  Factorisation of the Hilbert Space for Two-Sided Black Holes in Higher
  Dimensions}},}\ }\href@noop {} {\  (\bibinfo {year} {2024})},\ \Eprint
  {http://arxiv.org/abs/2410.23886} {arXiv:2410.23886 [hep-th]} \BibitemShut
  {NoStop}%
\bibitem [{\citenamefont {Magan}\ and\ \citenamefont
  {Wu}(2024)}]{Magan:2024nkr}%
  \BibitemOpen
  \bibfield  {author} {\bibinfo {author} {\bibfnamefont {Javier~M.}\
  \bibnamefont {Magan}}\ and\ \bibinfo {author} {\bibfnamefont {Qingyue}\
  \bibnamefont {Wu}},\ }\bibfield  {title} {\enquote {\bibinfo {title} {{Two
  types of quantum chaos: testing the limits of the Bohigas-Giannoni-Schmit
  conjecture}},}\ }\href@noop {} {\  (\bibinfo {year} {2024})},\ \Eprint
  {http://arxiv.org/abs/2411.08186} {arXiv:2411.08186 [quant-ph]} \BibitemShut
  {NoStop}%
\bibitem [{\citenamefont {Roos}(1970)}]{Roos:1970fm}%
  \BibitemOpen
  \bibfield  {author} {\bibinfo {author} {\bibfnamefont {H.}~\bibnamefont
  {Roos}},\ }\bibfield  {title} {\enquote {\bibinfo {title} {{Independence of
  local algebras in quantum field theory}},}\ }\href {\doibase
  10.1007/BF01646790} {\bibfield  {journal} {\bibinfo  {journal} {Commun. Math.
  Phys.}\ }\textbf {\bibinfo {volume} {16}},\ \bibinfo {pages} {238--246}
  (\bibinfo {year} {1970})}\BibitemShut {NoStop}%
\bibitem [{\citenamefont {Buchholz}(1974)}]{Buchholz:1973bk}%
  \BibitemOpen
  \bibfield  {author} {\bibinfo {author} {\bibfnamefont {Detlev}\ \bibnamefont
  {Buchholz}},\ }\bibfield  {title} {\enquote {\bibinfo {title} {{PRODUCT
  STATES FOR LOCAL ALGEBRAS}},}\ }\href {\doibase 10.1007/BF01646201}
  {\bibfield  {journal} {\bibinfo  {journal} {Commun. Math. Phys.}\ }\textbf
  {\bibinfo {volume} {36}},\ \bibinfo {pages} {287--304} (\bibinfo {year}
  {1974})}\BibitemShut {NoStop}%
\bibitem [{\citenamefont {Fewster}(2016)}]{fewster2016split}%
  \BibitemOpen
  \bibfield  {author} {\bibinfo {author} {\bibfnamefont {Christopher~J}\
  \bibnamefont {Fewster}},\ }\bibfield  {title} {\enquote {\bibinfo {title}
  {The split property for quantum field theories in flat and curved
  spacetimes},}\ }in\ \href@noop {} {\emph {\bibinfo {booktitle} {Abhandlungen
  aus dem Mathematischen Seminar der Universit{\"a}t Hamburg}}},\ Vol.~\bibinfo
  {volume} {86}\ (\bibinfo {organization} {Springer},\ \bibinfo {year} {2016})\
  pp.\ \bibinfo {pages} {153--175}\BibitemShut {NoStop}%
\bibitem [{\citenamefont {Gibbons}\ and\ \citenamefont
  {Hawking}(1977)}]{PhysRevD.15.2752}%
  \BibitemOpen
  \bibfield  {author} {\bibinfo {author} {\bibfnamefont {G.~W.}\ \bibnamefont
  {Gibbons}}\ and\ \bibinfo {author} {\bibfnamefont {S.~W.}\ \bibnamefont
  {Hawking}},\ }\bibfield  {title} {\enquote {\bibinfo {title} {Action
  integrals and partition functions in quantum gravity},}\ }\href {\doibase
  10.1103/PhysRevD.15.2752} {\bibfield  {journal} {\bibinfo  {journal} {Phys.
  Rev. D}\ }\textbf {\bibinfo {volume} {15}},\ \bibinfo {pages} {2752--2756}
  (\bibinfo {year} {1977})}\BibitemShut {NoStop}%
\bibitem [{\citenamefont {Raju}(2022)}]{Raju:2021lwh}%
  \BibitemOpen
  \bibfield  {author} {\bibinfo {author} {\bibfnamefont {Suvrat}\ \bibnamefont
  {Raju}},\ }\bibfield  {title} {\enquote {\bibinfo {title} {{Failure of the
  split property in gravity and the information paradox}},}\ }\href {\doibase
  10.1088/1361-6382/ac482b} {\bibfield  {journal} {\bibinfo  {journal} {Class.
  Quant. Grav.}\ }\textbf {\bibinfo {volume} {39}},\ \bibinfo {pages} {064002}
  (\bibinfo {year} {2022})},\ \Eprint {http://arxiv.org/abs/2110.05470}
  {arXiv:2110.05470 [hep-th]} \BibitemShut {NoStop}%
\bibitem [{Note3()}]{Note3}%
  \BibitemOpen
  \bibinfo {note} {In general the microstate of a thermodynamical system is
  defined as a pure state $\left |\psi \right \rangle $, which is
  indistinguishable from the equilibrium state represented by a thermal density
  matrix $\rho $, when probed by simple operators \cite
  {Climent:2024trz,Geng:2024jmm}, in the sense that \begin {equation*} \left
  \langle \psi \right |\protect \EuScript {O}_1...\protect \EuScript {O}_n\left
  |\psi \right \rangle =\protect \mathrm {Tr}(\rho \protect \tmspace
  +\thinmuskip {.1667em}\protect \EuScript {O}_1...\protect \EuScript
  {O}_n)\protect \tmspace +\thinmuskip {.1667em}. \label
  {eq:microstate_condition} \end {equation*}}\BibitemShut {NoStop}%
\bibitem [{Note7()}]{Note7}%
  \BibitemOpen
  \bibinfo {note} {In the present example of an eternal black hole, the phases
  arising from the global time evolution can be measured if two observers in a
  gedanken experimental set-up perform a non-local measurement by jumping into
  the black hole from different sides. They would detect a time-shift while
  comparing their respective times at their meeting point in the bulk, provided
  their watches were initially synchronized at asymptotic infinity \cite
  {Verlinde:2020upt}.}\BibitemShut {Stop}%
\bibitem [{\citenamefont {Verlinde}(2021)}]{Verlinde:2021kgt}%
  \BibitemOpen
  \bibfield  {author} {\bibinfo {author} {\bibfnamefont {Herman}\ \bibnamefont
  {Verlinde}},\ }\bibfield  {title} {\enquote {\bibinfo {title} {{Wormholes in
  Quantum Mechanics}},}\ }\href@noop {} {\  (\bibinfo {year} {2021})},\ \Eprint
  {http://arxiv.org/abs/2105.02129} {arXiv:2105.02129 [hep-th]} \BibitemShut
  {NoStop}%
\bibitem [{Note2()}]{Note2}%
  \BibitemOpen
  \bibinfo {note} {We refer to the semiclassical description, together with
  perturbative corrections as the effective description, while the full quantum
  gravity theory is the fundamental description, including non-perturbative
  corrections. The fundamental description is also sometimes termed as the
  boundary description.}\BibitemShut {Stop}%
\bibitem [{\citenamefont {Chandrasekaran}\ \emph {et~al.}(2022)\citenamefont
  {Chandrasekaran}, \citenamefont {Penington},\ and\ \citenamefont
  {Witten}}]{Chandrasekaran:2022eqq}%
  \BibitemOpen
  \bibfield  {author} {\bibinfo {author} {\bibfnamefont {Venkatesa}\
  \bibnamefont {Chandrasekaran}}, \bibinfo {author} {\bibfnamefont {Geoff}\
  \bibnamefont {Penington}}, \ and\ \bibinfo {author} {\bibfnamefont {Edward}\
  \bibnamefont {Witten}},\ }\bibfield  {title} {\enquote {\bibinfo {title}
  {{Large N algebras and generalized entropy}},}\ }\href@noop {} {\  (\bibinfo
  {year} {2022})},\ \Eprint {http://arxiv.org/abs/2209.10454} {arXiv:2209.10454
  [hep-th]} \BibitemShut {NoStop}%
\bibitem [{Note5()}]{Note5}%
  \BibitemOpen
  \bibinfo {note} {The rank of the Gram matrix may in principle directly be
  determined by either performing a Gram-Schmidt orthogonalization procedure,
  or equivalently by diagonalizing $G$. Since we have to find $d_\Omega $ in
  the limit $\Omega \to \infty $, this requires diagonalizing a matrix of
  infinite size, which is very hard. The resolvent method provides a simpler
  way to compute the eigenvalue density of the Gram matrix.}\BibitemShut
  {Stop}%
\bibitem [{Note4()}]{Note4}%
  \BibitemOpen
  \bibinfo {note} {This method was also used to give a microscopic
  interpretation of Bekenstein-Hawking entropy by considering a different
  family of microstates \cite
  {Balasubramanian:2022gmo,Balasubramanian:2022lnw,Climent:2024trz,Geng:2024jmm}}\BibitemShut
  {NoStop}%
\bibitem [{\citenamefont {Cvitanović}(1981)}]{CVITANOVIC198149}%
  \BibitemOpen
  \bibfield  {author} {\bibinfo {author} {\bibfnamefont {Predrag}\ \bibnamefont
  {Cvitanović}},\ }\bibfield  {title} {\enquote {\bibinfo {title} {Planar
  perturbation expansion},}\ }\href {\doibase
  https://doi.org/10.1016/0370-2693(81)90801-7} {\bibfield  {journal} {\bibinfo
   {journal} {Physics Letters B}\ }\textbf {\bibinfo {volume} {99}},\ \bibinfo
  {pages} {49--52} (\bibinfo {year} {1981})}\BibitemShut {NoStop}%
\bibitem [{\citenamefont {{Speicher}}(2009)}]{2009arXiv0911.0087S}%
  \BibitemOpen
  \bibfield  {author} {\bibinfo {author} {\bibfnamefont {Roland}\ \bibnamefont
  {{Speicher}}},\ }\bibfield  {title} {\enquote {\bibinfo {title} {{Free
  Probability Theory}},}\ }\href {\doibase 10.48550/arXiv.0911.0087} {\bibfield
   {journal} {\bibinfo  {journal} {arXiv e-prints}\ ,\ \bibinfo {eid}
  {arXiv:0911.0087}} (\bibinfo {year} {2009})},\ \Eprint
  {http://arxiv.org/abs/0911.0087} {arXiv:0911.0087 [math.PR]} \BibitemShut
  {NoStop}%
\bibitem [{\citenamefont {Bousso}(2002)}]{Bousso:2002ju}%
  \BibitemOpen
  \bibfield  {author} {\bibinfo {author} {\bibfnamefont {Raphael}\ \bibnamefont
  {Bousso}},\ }\bibfield  {title} {\enquote {\bibinfo {title} {{The Holographic
  principle}},}\ }\href {\doibase 10.1103/RevModPhys.74.825} {\bibfield
  {journal} {\bibinfo  {journal} {Rev. Mod. Phys.}\ }\textbf {\bibinfo {volume}
  {74}},\ \bibinfo {pages} {825--874} (\bibinfo {year} {2002})},\ \Eprint
  {http://arxiv.org/abs/hep-th/0203101} {arXiv:hep-th/0203101} \BibitemShut
  {NoStop}%
\bibitem [{\citenamefont {Furuya}\ \emph {et~al.}(2023)\citenamefont {Furuya},
  \citenamefont {Lashkari}, \citenamefont {Moosa},\ and\ \citenamefont
  {Ouseph}}]{Furuya:2023fei}%
  \BibitemOpen
  \bibfield  {author} {\bibinfo {author} {\bibfnamefont {Keiichiro}\
  \bibnamefont {Furuya}}, \bibinfo {author} {\bibfnamefont {Nima}\ \bibnamefont
  {Lashkari}}, \bibinfo {author} {\bibfnamefont {Mudassir}\ \bibnamefont
  {Moosa}}, \ and\ \bibinfo {author} {\bibfnamefont {Shoy}\ \bibnamefont
  {Ouseph}},\ }\bibfield  {title} {\enquote {\bibinfo {title} {{Information
  loss, mixing and emergent type III$_1$ factors}},}\ }\href@noop {} {\
  (\bibinfo {year} {2023})},\ \Eprint {http://arxiv.org/abs/2305.16028}
  {arXiv:2305.16028 [hep-th]} \BibitemShut {NoStop}%
\bibitem [{\citenamefont {Witten}(2021)}]{Witten:2021jzq}%
  \BibitemOpen
  \bibfield  {author} {\bibinfo {author} {\bibfnamefont {Edward}\ \bibnamefont
  {Witten}},\ }\bibfield  {title} {\enquote {\bibinfo {title} {{Why Does
  Quantum Field Theory In Curved Spacetime Make Sense? And What Happens To The
  Algebra of Observables In The Thermodynamic Limit?}}}\ }\href@noop {} {\
  (\bibinfo {year} {2021})},\ \Eprint {http://arxiv.org/abs/2112.11614}
  {arXiv:2112.11614 [hep-th]} \BibitemShut {NoStop}%
\bibitem [{\citenamefont {Leutheusser}\ and\ \citenamefont
  {Liu}(2021{\natexlab{a}})}]{Leutheusser:2021qhd}%
  \BibitemOpen
  \bibfield  {author} {\bibinfo {author} {\bibfnamefont {Samuel}\ \bibnamefont
  {Leutheusser}}\ and\ \bibinfo {author} {\bibfnamefont {Hong}\ \bibnamefont
  {Liu}},\ }\bibfield  {title} {\enquote {\bibinfo {title} {{Causal
  connectability between quantum systems and the black hole interior in
  holographic duality}},}\ }\href@noop {} {\  (\bibinfo {year}
  {2021}{\natexlab{a}})},\ \Eprint {http://arxiv.org/abs/2110.05497}
  {arXiv:2110.05497 [hep-th]} \BibitemShut {NoStop}%
\bibitem [{\citenamefont {Leutheusser}\ and\ \citenamefont
  {Liu}(2021{\natexlab{b}})}]{Leutheusser:2021frk}%
  \BibitemOpen
  \bibfield  {author} {\bibinfo {author} {\bibfnamefont {Samuel}\ \bibnamefont
  {Leutheusser}}\ and\ \bibinfo {author} {\bibfnamefont {Hong}\ \bibnamefont
  {Liu}},\ }\bibfield  {title} {\enquote {\bibinfo {title} {{Emergent times in
  holographic duality}},}\ }\href@noop {} {\  (\bibinfo {year}
  {2021}{\natexlab{b}})},\ \Eprint {http://arxiv.org/abs/2112.12156}
  {arXiv:2112.12156 [hep-th]} \BibitemShut {NoStop}%
\bibitem [{Note6()}]{Note6}%
  \BibitemOpen
  \bibinfo {note} {A similar structure of the boundary Hilbert space was
  studied \cite {Leutheusser:2022bgi}, where general states with a gravity dual
  were considered.}\BibitemShut {Stop}%
\bibitem [{\citenamefont {Hartle}\ and\ \citenamefont
  {Hawking}(1976)}]{PhysRevD.13.2188}%
  \BibitemOpen
  \bibfield  {author} {\bibinfo {author} {\bibfnamefont {J.~B.}\ \bibnamefont
  {Hartle}}\ and\ \bibinfo {author} {\bibfnamefont {S.~W.}\ \bibnamefont
  {Hawking}},\ }\bibfield  {title} {\enquote {\bibinfo {title} {Path-integral
  derivation of black-hole radiance},}\ }\href {\doibase
  10.1103/PhysRevD.13.2188} {\bibfield  {journal} {\bibinfo  {journal} {Phys.
  Rev. D}\ }\textbf {\bibinfo {volume} {13}},\ \bibinfo {pages} {2188--2203}
  (\bibinfo {year} {1976})}\BibitemShut {NoStop}%
\bibitem [{\citenamefont {Harlow}(2016)}]{Harlow:2014yka}%
  \BibitemOpen
  \bibfield  {author} {\bibinfo {author} {\bibfnamefont {Daniel}\ \bibnamefont
  {Harlow}},\ }\bibfield  {title} {\enquote {\bibinfo {title} {{Jerusalem
  Lectures on Black Holes and Quantum Information}},}\ }\href {\doibase
  10.1103/RevModPhys.88.015002} {\bibfield  {journal} {\bibinfo  {journal}
  {Rev. Mod. Phys.}\ }\textbf {\bibinfo {volume} {88}},\ \bibinfo {pages}
  {015002} (\bibinfo {year} {2016})},\ \Eprint {http://arxiv.org/abs/1409.1231}
  {arXiv:1409.1231 [hep-th]} \BibitemShut {NoStop}%
\bibitem [{\citenamefont {Eynard}\ \emph {et~al.}(2015)\citenamefont {Eynard},
  \citenamefont {Kimura},\ and\ \citenamefont {Ribault}}]{Eynard2015}%
  \BibitemOpen
  \bibfield  {author} {\bibinfo {author} {\bibfnamefont {Bertrand}\
  \bibnamefont {Eynard}}, \bibinfo {author} {\bibfnamefont {Taro}\ \bibnamefont
  {Kimura}}, \ and\ \bibinfo {author} {\bibfnamefont {Sylvain}\ \bibnamefont
  {Ribault}},\ }\bibfield  {title} {\enquote {\bibinfo {title} {Random
  matrices},}\ }\href@noop {} {\  (\bibinfo {year} {2015})},\ \Eprint
  {http://arxiv.org/abs/1510.04430} {arXiv:1510.04430 [math-ph]} \BibitemShut
  {NoStop}%
\end{thebibliography}%

\end{document}